\documentclass[aps,prc]{revtex4}
\usepackage{epsfig}
\newcommand{\beq}{\begin{equation}}
\newcommand{\eeq}{\end{equation}}
\newcommand{\bea}{\begin{eqnarray}}
\newcommand{\eea}{\end{eqnarray}}
\newcommand{\bef}{\begin{figure}}
\newcommand{\eef}{\end{figure}}
\newcommand{\bce}{\begin{center}}
\newcommand{\ece}{\end{center}}
\newcommand{\eg}{{\it e.g.}}
\newcommand{\ie}{{\it i.e.}}
\newcommand{\etal}{{\it et al.}}
\def\lsim{\mathrel{\rlap{\lower4pt\hbox{\hskip1pt$\sim$}}
    \raise1pt\hbox{$<$}}}         %less than or approx. symbol
\def\gsim{\mathrel{\rlap{\lower4pt\hbox{\hskip1pt$\sim$}}
    \raise1pt\hbox{$>$}}}         %greater than or approx. symbol
\begin{document}
%\draft
%\tightenlines
%
%
\title
{Hadronic Production of Thermal Photons}
 
\author
{Simon Turbide$^1$, Ralf Rapp$^{2}$, and  Charles Gale$^1$}
 
\affiliation 
{$^1$ Department of Physics, McGill University, 3600 University Street, 
    Montreal, Canada H3A 2T8\\
 $^2$ NORDITA, Blegdamsvej 17, DK-2100 Copenhagen, Denmark}

\date{\today}

\begin{abstract}
We study the thermal emission of photons from hot and dense strongly 
interacting hadronic matter at temperatures close to the expected phase 
transition to the Quark-Gluon Plasma (QGP). Earlier calculations of 
photon radiation from ensembles of interacting mesons are re-examined 
with additional constraints, including new production channels as well 
as an assessment of hadronic form factor effects. Whereas 
strangeness-induced photon yields turn out to be moderate, the hitherto 
not considered $t$-channel exchange of $\omega$-mesons is found to 
contribute appreciably for photon energies above $\sim$~1.5~GeV. The 
role of baryonic effects is assessed using existing many-body 
calculations of lepton pair production. We argue that our combined 
results constitute a rather realistic emission rate, appropriate for 
applications in relativistic heavy-ion collisions. Supplemented with 
recent evaluations of QGP emission, and an estimate for primordial 
(hard) production, we compute photon spectra at SPS, RHIC and LHC 
energies.
\end{abstract}

\maketitle

%\pacs{}

%%%%%%%%%%%%%%%%%%%%%%%%%%%%%%%%%%%%%%%%%%%%%%%%%%%%%%%%%%%%%%%%%%%%%%%%
\section{Introduction}
%%%%%%%%%%%%%%%%%%%%%%%%%%%%%%%%%%%%%%%%%%%%%%%%%%%%%%%%%%%%%%%%%%%%%%%%
Electromagnetic radiation (real and virtual photons) from central 
high-energy collisions of heavy nuclei ($A$-$A$) has the potential to 
directly probe the high temperature and high density phases of these 
reactions. This is so because photons and dileptons suffer little final 
state interaction, and also because their emission rate is a rapidly 
varying function of local intensive variables. It is therefore hoped 
that electromagnetic (e.m.) observables reveal novel features of 
strongly interacting systems under extreme conditions, including the 
quark-gluon plasma (QGP)~\cite{qgp}, which is but one facet of a rich 
many-body problem. 

More than a decade of careful experiments and of their theoretical
interpretation has made clear that a proof of the existence of the QGP 
requires analyses of many complementary measurements.
This, in particular, applies 
to thermal production of photons and dileptons, since their      
emissivity is encoded in the same theoretical quantity, \ie~the 
e.m. current-current correlation function in   
strong-interaction matter~\cite{formal}.

In this article we focus on the study of real photon emission (see
Ref.~\cite{PT02} for recent reviews). 
An identification of the QGP component in the spectra requires a 
reliable assessment of competing sources, most notably primordial 
production in hard nucleon-nucleon ($N$-$N$) collisions~\cite{Aur99}, 
as well as the later hadronic emission.  Owing to the  
vanishing invariant mass of real photons, all three sources
generate essentially smooth spectra, measured as a function
of transverse momentum, $q_t$.
The key properties are then their strength together with a typical 
spectral slope. 
For thermal radiation the latter is governed by temperature (as well 
as collective matter expansion), whereas the overall normalisation is 
set by a combination of microscopic production mechanisms and the 
space-time volume of the interacting matter.
Naively, one expects QGP radiation (from large temperatures and small 
volumes) to dominate at large $q_t$, and hadron gas (HG) emission at 
small $q_t$. In practice, \ie~within a heavy-ion environment, both ends 
of the thermal spectra are masked: there is a large background at 
low $q_t$, and the primordial component from hard $N$-$N$ collisions 
feeds the high $q_t$. The primordial yield in $A$-$A$ collisions is 
expected to be modified (as compared to the free $N$-$N$ case) by 
nuclear transverse-momentum ($k_T$-) broadening~\cite{cronin,kt} and 
shadowing~\cite{shadow}. In general, the strength of thermal 
emissivities in both hadron gas and 
QGP can also be substantially affected by microscopic in-medium 
effects. Furthermore, collective (transverse) expansion induces a 
spectral hardening being more pronounced in the later (hadronic) 
phases. 

Most of the above issues have been addressed in various forms
before, \eg~within hydrodynamic simulations~\cite{Dumi95,SS01,Huo02},
transport/cascade models~\cite{Hal98,SG98,hbek02} or simple fireball 
models~\cite{RW00-2,GKP00} (where consistency with dilepton observables
has been emphasised).     
In the present paper we are mostly concerned with hadron gas 
emission, improving previous analyses in several respects.  
$\pi\rho a_1$-meson gas rates are revisited and extended, including 
strangeness reactions, heavy resonances as well as $t$-channel 
exchange of the $\omega(782)$. The effects of baryons -- which are 
known to be of prime importance in the dilepton context~\cite{RW00} 
(as recently further supported by measurements at lower SPS 
energy~\cite{ceres40}) -- are extracted based on the same 
spectral densities that successfully describe the dilepton data,
thus ensuring consistency. This also holds for our modelling of the 
space-time evolution of heavy-ion reactions, which, in particular,
accounts for effects of chemical off-equilibrium in the hadronic phase. 
Finally, for comparison 
with data from SPS (and for predictions at RHIC 
and LHC), hadronic emission is supplemented with an estimate of hard 
(primordial) production, an assessment of the Cronin effect,  as well as 
QGP radiation (complete to leading order) 
within a schematic evolution of the collision dynamics.  
 
The article is organised as follows. In Sec.~\ref{sec_had} we briefly recall 
the definition of emission rates, followed by the evaluation of hadronic 
photon radiation from 
(i) a hot meson gas using a massive Yang-Mills (MYM) approach, extended  
to the $SU(3)$ (strangeness) sector, with a quantitative assessment of hadronic 
form factors;
(ii) a baryon-rich environment, as well as, (iii) additional meson sources 
beyond the MYM framework. 
The total hadronic rate is then compared with a recent
leading-order QGP result. 
In Sec.~\ref{sec_urhic} we employ a simple dynamical model to compare
integrated yields with experimental measurements at SPS, and to provide 
predictions for RHIC and LHC. 
Finally, Sec.~\ref{sec_concl} contains a summary and conclusions.

%%%%%%%%%%%%%%%%%%%%%%%%%%%%%%%%%%%%%%%%%%%%%%%%%%%%%%%%%%%%%%%%%%%%%%%%%%%%
\section{Thermal Photon Rates from Hadronic Matter} 
\label{sec_had}
%%%%%%%%%%%%%%%%%%%%%%%%%%%%%%%%%%%%%%%%%%%%%%%%%%%%%%%%%%%%%%%%%%%%%%%%%%%%

%%%%%%%%%%%%%%%%%%%%%%%%%%%%%%%%%%%%%%%%%%%%%%%%%%%%%%%%%%%%%%%%%%%%%%%%%%%%
\subsection{Definitions}                    
\label{sec_def} 
%%%%%%%%%%%%%%%%%%%%%%%%%%%%%%%%%%%%%%%%%%%%%%%%%%%%%%%%%%%%%%%%%%%%%%%%%%%%
In this work we will adopt two different approaches to calculate 
thermal photon production from hot and dense matter, taking advantage
of the respective strengths of each one. 

Within the thermal field theory framework, the differential photon emission rate  
from an equilibrated system can be written in compact form as~\cite{formal} 
\beq
\frac{dR_\gamma}{d^4q} = P_{\mu\nu} \ W^{\mu\nu} \  
\label{rate4}
\eeq
where 
\beq
P_{\mu\nu}=4\pi\alpha \int \frac{d^3p}{(2\pi)^3 2p_0} \sum\limits_\lambda
           \varepsilon_\mu^*(\lambda,p) \ \varepsilon_\nu(\lambda,p) \ 
           \delta^{(4)}(p-q) 
\eeq
denotes the (polarisation-summed) photon tensor to lowest order in the 
e.m.~coupling $\alpha$. The key quantity is the ha\-dronic tensor 
$W^{\mu\nu}(q;\mu_b,T)$, which contains the effects of strong 
interactions to all orders and is directly related to the imaginary 
part of the in-medium photon self-energy (or e.m.~current-current 
correlation function), $\Pi_{\rm em}$, via 
\beq
W^{\mu\nu}=(-2) \ f^B(q_0;T) \ {\rm Im} \Pi_{\rm em}^{\mu\nu}
\eeq 
with $f^B(q_0;T)=1/(\exp[q_0/T]-1)$ the Bose distribution function 
(note the similarity of Eq.~(\ref{rate4}) to the dilepton rate,
which is obtained upon replacing $P_{\mu\nu}$ with the lepton
tensor $L_{\mu\nu}$). The photon rate can therefore be cast into the 
form
\beq
q_0\frac{dR_\gamma}{d^3q} = -\frac{\alpha}{\pi^2} \ f^B(q_0;T) \
        {\rm Im}\Pi_{\rm em}^T(q_0=q;T) \ .   
\label{rate}
\eeq 
Invoking the vector dominance model (VDM) \cite{vdm}, $\Pi_{\rm em}$ can be directly
related to in-medium vector-meson spectral functions (see below), 
which makes Eq.~(\ref{rate}) particularly suitable for non-perturbative model
calculations at low and moderate energies and momenta. 

Alternatively, the photon emission can also be computed from relativistic 
kinetic theory. For a given reaction $1+2\rightarrow 3+\gamma$, the 
pertinent rate becomes 
\begin{equation}
q_0 \frac{dR_\gamma}{d^3q} =\int \frac{d^3p_1}{2(2\pi)^3E_1}
\frac{d^3p_2}{2(2\pi)^3E_2}\frac{d^3p_3}{2(2\pi)^3E_3}(2\pi)^4
\delta^{(4)}(p_1+p_2\rightarrow p_3+q)
\left|{\cal M}\right|^2\frac{f(E_1)f(E_2)[1\pm f(E_3)]}{2(2\pi)^3}
\label{rate_kin}
\end{equation} 
where ${\cal M}$ is the invariant scattering matrix element. For photon 
self-energies of no more than two-loop order, its imaginary
part reduces to tree level diagrams, in which case it is usually more 
convenient to find thermal rates with the kinetic formula, 
Eq.~(\ref{rate_kin}). The perturbative character of this approach thus 
makes it advantageous over Eq.~(\ref{rate}) at moderate and 
high energies and momenta.

%%%%%%%%%%%%%%%%%%%%%%%%%%%%%%%%%%%%%%%%%%%%%%%%%%%%%%%%%%%%%%%%%%%%%%%
\subsection{Meson Gas and Hadronic Form Factors}
\label{sec_mesgas}
%%%%%%%%%%%%%%%%%%%%%%%%%%%%%%%%%%%%%%%%%%%%%%%%%%%%%%%%%%%%%%%%%%%%%%%
To describe photon-producing reactions in a gas  
consisting of light pseudo-scalar, vector and axial vector 
mesons ($\pi, K, \rho, K^*, a_1$) we employ  
the massive Yang-Mills (MYM) approach, which is 
capable of yielding adequate hadronic phenomenology at tree
level with a rather limited set of adjustable parameters.  Vector and 
axial vector fields are implemented into an effective
nonlinear $\sigma$-model Lagrangian as massive gauge fields of 
the chiral U$(3)_L \times$ U$(3)_R$ symmetry\cite{gomm,Song93}:
\bea
{\cal L} &=& \textstyle{\frac{1}{8}} F_\pi^2 {\rm Tr} D_\mu U D^\mu U^\dag + \frac{1}{8}
F_\pi^2 {\rm Tr} M (U + U^\dag -2)\nonumber \\& &  - \textstyle{\frac{1}{2}}
{\rm Tr} \left(F_{\mu \nu}^L
{F^L}^{\mu \nu} + F_{\mu \nu}^R {F^R}^{\mu \nu} \right) + m_0^2 {\rm Tr}
\left(A_\mu^L {A^L}^{\mu \nu} + A_\mu^R {A^R}^\mu\right)+
\gamma {\rm Tr}F_{\mu \nu}^L U F^{R \mu \nu}U^\dag\nonumber 
\\& & -i\xi {\rm Tr}\left(D_\mu UD_\nu U^\dag F^{L \mu \nu}
+D_\mu U^\dag D_\nu U F^{R \mu \nu}\right)\ .
\label{Lmym}
\eea
In the above,
\bea
\lefteqn{U = \exp \left( \frac{2 i}{F_\pi} \sum_i \frac{\phi_i
\lambda_i}{\sqrt{2}}\right) = \exp\left( \frac{2 i}{F_\pi} 
\mbox{\boldmath $\phi$} \right)\ ,}\ \nonumber \\
 & & A_\mu^L = \textstyle{\frac{1}{2}}(V_\mu + A_\mu)\ , \nonumber \\
 & & A_\mu^R = \textstyle{\frac{1}{2}}(V_\mu - A_\mu)\ , \nonumber \\
 & & F_{\mu \nu}^{L, R}  = \partial_\mu A_\nu^{L, R} - \partial_\nu A_{\mu}^{L, R} -
i g_0 \left[A_{\mu}^{L, R}, A_\nu^{L, R} \right]\ ,\nonumber \\
 & & D_\mu U = \partial_\mu U - i g_0 A_\mu^L U + i g_0 U A_\mu^R\ ,\nonumber \\
 & & M = \frac{2}{3} \left[ m_K^2 + \frac{1}{2} m_\pi^2\right] - \frac{2}{\sqrt{3}}
(m_K^2 - m_\pi^2) \lambda_8\ .
\eea
Note that $F_\pi$ = 135 MeV and that $\lambda_i$ is a Gell-Mann matrix.  
$\phi, V_\mu$ and $A_\mu$ are, respectively, the pseudo-scalar, vector 
and axial vector meson matrices. Note that this form of the 
interaction enables a coherent treatment of the strange and nonstrange
fields.

Let us first concentrate on the nonstrange sector ($\pi\rho a_1$). 
Following Ref.~\cite{Song93}, the four unspecified parameters in the 
Lagrangian can be inferred from the masses and widths of the $\rho$ 
and $a_1$, allowing for two possible solutions:
\bea
({\rm I}):\ \ \ \ \ \tilde{g} = 10.3063, \ \gamma = 0.3405,\ 
\xi = 0.4473,\ m_0 = 0.6253\ \mbox{GeV}\ ;\nonumber \\
({\rm II}):\ \ \ \ \,
\tilde{g} = 6.4483,\ \gamma = - 0.2913,\ \xi=0.0585,\ m_0 = 0.875\ \mbox{GeV}\ ,
\eea
where $\tilde{g} = g_0 / \sqrt{1 - \gamma}$. 
In the absence of additional empirical constraints, the use of one set over
another is difficult to justify. However, in Ref.~\cite{gao} it has been 
suggested to invoke the experimental determination of $D$- to $S$-wave content 
in the final state of the $a_1\to\rho\pi$ decay. For the two 
parameter sets given above one finds
\bea
&({\rm I}): &  D/S = 0.36
\nonumber \\
&({\rm II}):&  D/S = - 0.099  \  ,  
\eea 
to be compared to the experimental value~\cite{PDG} $- 0.107 \pm$ 0.016. 
From here on, we therefore employ parameter set II.

We proceed with a systematic evaluation of all processes generating
photons based on the interaction vertices contained in 
Eq.~(\ref{Lmym}).  The explicit reactions considered include all 
possible $s$-, $t$- and $u$-channel (Born-) graphs for the reactions:
$X+Y\rightarrow Z+\gamma$, $\rho\rightarrow Y+Z+\gamma$ and 
$K^*\rightarrow Y+Z+\gamma$. For $X$, $Y$, $Z$ we have each combination 
of $\rho, \pi,K^*,K$ mesons which respect the conservation of charge, 
isospin, strangeness and G-parity defined for non-strange mesons. The 
axial $a_1$ meson has been considered as exchange particle only (the 
$a_1\to \pi \gamma$ decay is automatically incorporated via $s$-channel 
$\pi\rho$ scattering).  Using Eq.~(\ref{rate_kin}) the thermal photon 
production rates are readily obtained from the coherently summed matrix
elements in each channel, and convenient parametrisations thereof 
are given in Appendix~\ref{para}. 
The parametrisations for the nonstrange reaction channels in the present 
work differ from the ones quoted in 
Ref.~\cite{SF98}, which are based on Ref.~\cite{Song93}, in two respects. 
First, in the latter article the amplitude in two channels, as
written, violates the Ward identity~\cite{songref}.  
Second, the choice of
parameters underlying Ref.~\cite{SF98} yields an $D/S$ ratio which is
at variance with the experimental value. This 
is corrected here, together with amplitudes which have been verified 
for current conservation. As in Ref.~\cite{Song93}, 
parameter set (II) yields smaller emission rates 
than set (I) for the (leading) $\pi\rho\to\pi\gamma$ process at high 
energies. In addition, our results using set (II) are another 
$\sim$40\% smaller than the corresponding ones in 
Ref.~\cite{Song93}. 

\begin{figure}
%\vspace{-1cm}
\bce
\epsfig{file=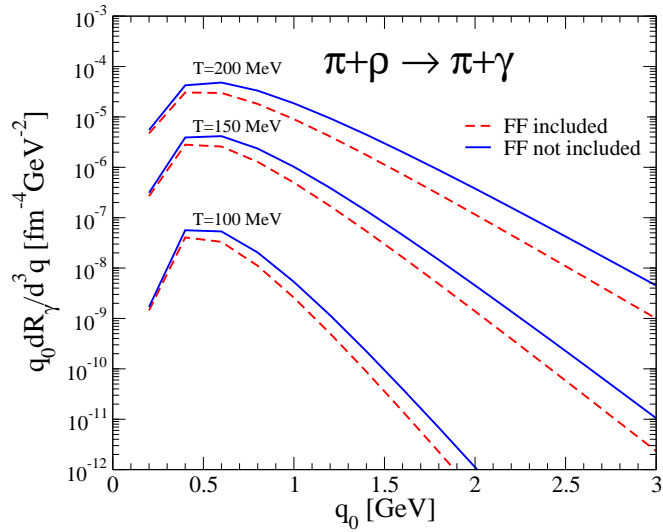,width=8.5cm,angle=-90}
\ece
\caption{(Color online) The effect of hadronic form factors on a typical 
photon-producing reaction in a nonstrange meson gas, 
$\pi \rho \to \pi \gamma$, for 3 different temperatures. Solid
and dashed lines are without and with the inclusion of $F(\bar t)^4$,
respectively.}
\label{fig_FF}
\end{figure}

An important element in applying effective hadronic models
at moderate and high momentum transfers is the use of hadronic
vertex form factors simulating finite-size effects, which are not
accounted for in the $\pi\rho a_1$ calculations of Ref.~\cite{Song93}.
For some of these reactions form factor effects have first been  
studied in Ref.~\cite{KLS91} and found to give a typical net
suppression over the bare graphs by an appreciable factor of $\sim$3 at
photon energies around  $q_0$$\simeq$2.5~GeV. We perform an estimate of
their effect in the present context as follows. To be consistent with 
the procedure adopted before for dilepton production~\cite{RG99}, we 
assume a standard dipole form for each hadronic vertex appearing in the 
amplitudes,
\beq
F(t)=\left(\frac{2\Lambda^2}{2\Lambda^2-t}\right)^2 \ . 
\label{ff_t}
\eeq
We then approximate the four-momentum transfer in a given $t$-channel
exchange of meson $X$ by its average $\bar t$ according to
\begin{equation}
\left(\frac{1}{m_{X}^2-\bar{t}}\right)^2
\left(\frac{2\Lambda^2}{2\Lambda^2-\bar{t}}\right)^8
=\frac{1}{4E^2}\int^{4E^2}_{0}
\frac{dt(2\Lambda^2)^8}{(m_{X}^2-t)^2(2\Lambda^2-t)^8} \ .
\label{ff_av}
\end{equation}
The averaging procedure allows to factorize
form factors and amplitudes which much facilitates the task of 
rendering the final expression gauge invariant, since we can then 
simply multiply the rate parametrisations with $F(\bar t)^4$ employing 
a typical hadronic scale, $\Lambda=1$~GeV~\cite{RG99}, for the cutoff 
parameter. The impact of the form factors is indicated by the 
difference of solid and dashed curves in Fig.~\ref{fig_FF}, which 
should still be considered as a conservative estimate of the 
suppression effect.  The reduction of the rate in the 2-3~GeV region 
amounts to a factor of $\sim$4, quite in line with the earlier 
estimates of Ref.~\cite{KLS91}.

%%%%%%%%%%%%%%%%%%%%%%%%%%%%%%%%%%%%%%%%%%%%%%%%%%%%%%%%%%%%%%%%%%%%%%%%
%\subsection{Mesonic Reactions with Strangeness}
%\label{sec_strange}
%%%%%%%%%%%%%%%%%%%%%%%%%%%%%%%%%%%%%%%%%%%%%%%%%%%%%%%%%%%%%%%%%%%%%%%%
Rather little attention has been paid in the literature so far to the 
calculation of photon emission rates involving strange particles, 
mostly because existing analyses have found them to be quantitatively 
suppressed.  In Ref.~\cite{hag94}, the channel $K_1 \to K \gamma$ was 
investigated and shown to be appreciable 
relative to non-strange sources only in a limited kinematical domain. 
Also, results obtained with $SU(3)$ chiral reduction 
formulae~\cite{lyz98}, coupled with an expansion in temperature, suggest 
that the strangeness contribution is not large. 
Here, we seek to quantify the latter relative to the $\pi\rho a_1$ 
emissivities within the same effective Lagrangian framework as encoded 
in the $SU(3)$ extension implicit in Eq.~(\ref{Lmym}). To optimally 
reproduce the  (measured) hadronic phenomenology, we are, however, lead 
to decouple the non-strange axial vector meson ($a_1$) from the 
strangeness sector. This allows to simultaneously satisfy the 
(electromagnetic) Ward identities and fix both the strange vector mass, 
$m_{K^*}$=895~MeV, independent of the $\rho$ mass, and the universal 
coupling constant as to match the empirical  value~\cite{PDG} of the 
$K^*$ width, $\Gamma (K^* \to K \pi )$$\simeq$50~MeV. 
\begin{figure}[!ht]
\bce
\epsfig{file=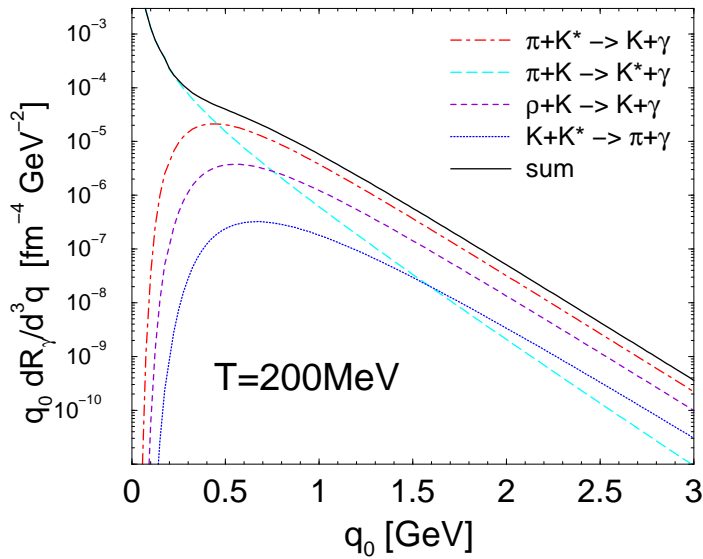,width=7.5cm,angle=-90}
\ece
\caption{(Color online) Photon-producing reaction rates involving strange mesons
at a temperature $T$=200~MeV with  form factor effects included.}
\label{fig_strange}
\end{figure}
The specific channels included are: 
$\pi K^* \to K \gamma$, $K \rho \to K \gamma$, $\pi K \to K^*
\gamma$, $K K^* \to \pi \gamma$, $K K \to \rho \gamma$, and 
$K^* \to \pi K \gamma$. Again, all possible isospin combinations are 
accounted for in the rate calculations, which have been parameterized in 
functional form in Appendix~\ref{para}. For a temperature of 200~MeV, 
the leading production channels are shown in Fig.~\ref{fig_strange}, 
with hadronic form factors implemented following the same procedure,
Eq.(\ref{ff_av}), as before (the last two contributions 
enumerated above have been  omitted, as they represent negligible 
increments).  At all energies of practical relevance ($q_0\ge0.5$~GeV)  
and including form factors, 
the $K^* \pi \to K\gamma$ reaction, mediated by $t$-channel 
$\pi$-exchange, turns out to be the main emission source, which is in 
complete analogy to the $\pi\rho\to\pi\gamma$ reaction in the 
nonstrange sector. In line with estimates in \cite{Ra02c}, the former 
constitutes $\sim$40\% of the latter around $q_0$=1~GeV, being reduced 
to about 20\% at $q_0$=3~GeV.

%%%%%%%%%%%%%%%%%%%%%%%%%%%%%%%%%%%%%%%%%%%%%%%%%%%%%%%%%%%%%%%%%%%%%%%
\subsection{In-Medium Selfenergies with Baryons}
\label{sec_self}
%%%%%%%%%%%%%%%%%%%%%%%%%%%%%%%%%%%%%%%%%%%%%%%%%%%%%%%%%%%%%%%%%%%%%%%
It is important to realize that thermal emission rates of dileptons 
and photons are intimately connected, both being based on the 
e.m.~current-current correlator, albeit evaluated in distinct 
kinematical domains, \ie~timelike ($M^2=q_0^2-q^2>0$) vs. lightlike 
($M^2=0$), respectively. The latter may imply the prevalence of 
different processes in the corresponding observable, but consistency 
can and should be tested.  In particular, we recall that baryonic 
sources are very important~\cite{RW00,Ra02a} for understanding the 
observed excess in low-mass ($M\le 1$~GeV) dilepton production in 
(semi-) central $Pb$-$Au$ collisions at both full 
(160~AGeV)~\cite{ceres160} and lower (40~AGeV)~\cite{ceres40} SPS 
energy. It is therefore mandatory to scrutinize the role of baryons 
in photon production, especially since most investigations thus far not 
revealed substantial contributions~\cite{SYZ97,LB98,Alam03}. 

We here make use of the hadronic many-body calculations of the in-medium
$\rho$(770) spectral function~\cite{RCW97,RUBW98,RW99}, which, when 
evaluated for $M^2\to 0$, directly yield pertinent photon emission 
rates via Eq.~(\ref{rate}). Within the vector dominance model (VDM), 
one has (schematically)
\beq
{\rm Im} \Pi_{\rm em} = \sum\limits_{V=\rho,\omega,\phi} \frac{m_V^4}{g_V^2} 
  {\rm Im}D_V
\eeq
($m_V$, $g_V$ and ${\rm Im}D_V$: vector-meson masses, coupling constants and
spectral functions, respectively). In the following we focus on contributions 
arising from the $\rho$-meson, which are dominant since 
$g_\rho^2/g_\omega^2$$ \simeq $10. In addition to in-medium effects in the 
pion cloud of the $\rho$-meson (encoded in a modified two-pion decay width, 
$\Sigma_{\rho\pi\pi}$), resonant $\rho$-$h$ ($h$=$\pi,K,\rho,N,\Delta$,...)
interactions are incorporated through self-energy expressions of type
\bea
\Sigma_{\rho h}^{\mu\nu}(q_0,\vec q;T)=\int \frac{d^3p}{(2\pi)^3}
\frac{1}{2\omega_h(p)} [f^h(\omega_h(p))-
f^{\rho h}(\omega_h(p)+q_0)] \ {\cal M}_{\rho h}^{\mu\nu}(p,q) \ ,
\label{sigmamunu}
\eea
where the isospin averaged $\rho$ scattering
amplitude ${\cal M}_{\rho h}$ is integrated over the thermal distribution
$f^h(\omega_h(p))=[\exp(\omega_h(p))/T\pm 1]^{-1}$  of the corresponding
hadron species $h$ with $\omega_h(p)=\sqrt{m_h^2+\vec p^2}$.
The advantage of writing the self-energy in terms of the forward scattering
amplitude is that in-medium resonance widths, accounting for higher order
effects in temperature and density, are readily implemented without facing
problems of double-counting. The latter becomes more difficult to keep
track of when evaluating higher order topologies in the kinetic-theory
approach represented by Eq.~(\ref{rate_kin})~\cite{majgale}. 
All of the resonances used in constructing the $\rho$ self-energy are enumerated
in Refs.~\cite{RG99,RW99}, which also contains more details on how the
interactions are constrained by hadronic phenomenology.

\begin{figure}[!ht]
%\vspace{-1cm}
\bce
\epsfig{file=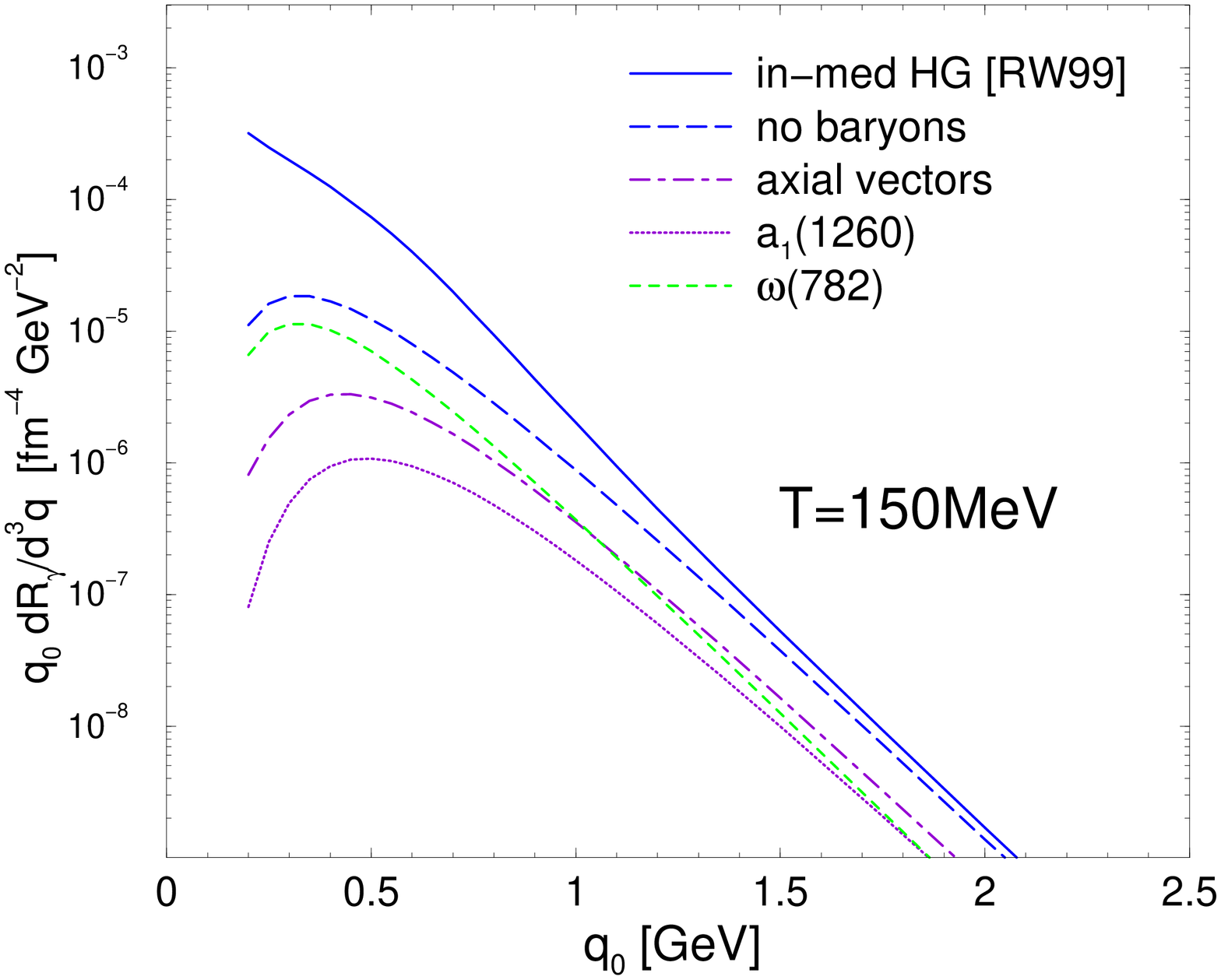,width=6.85cm,angle=-90}
\hspace{1cm}
\epsfig{file=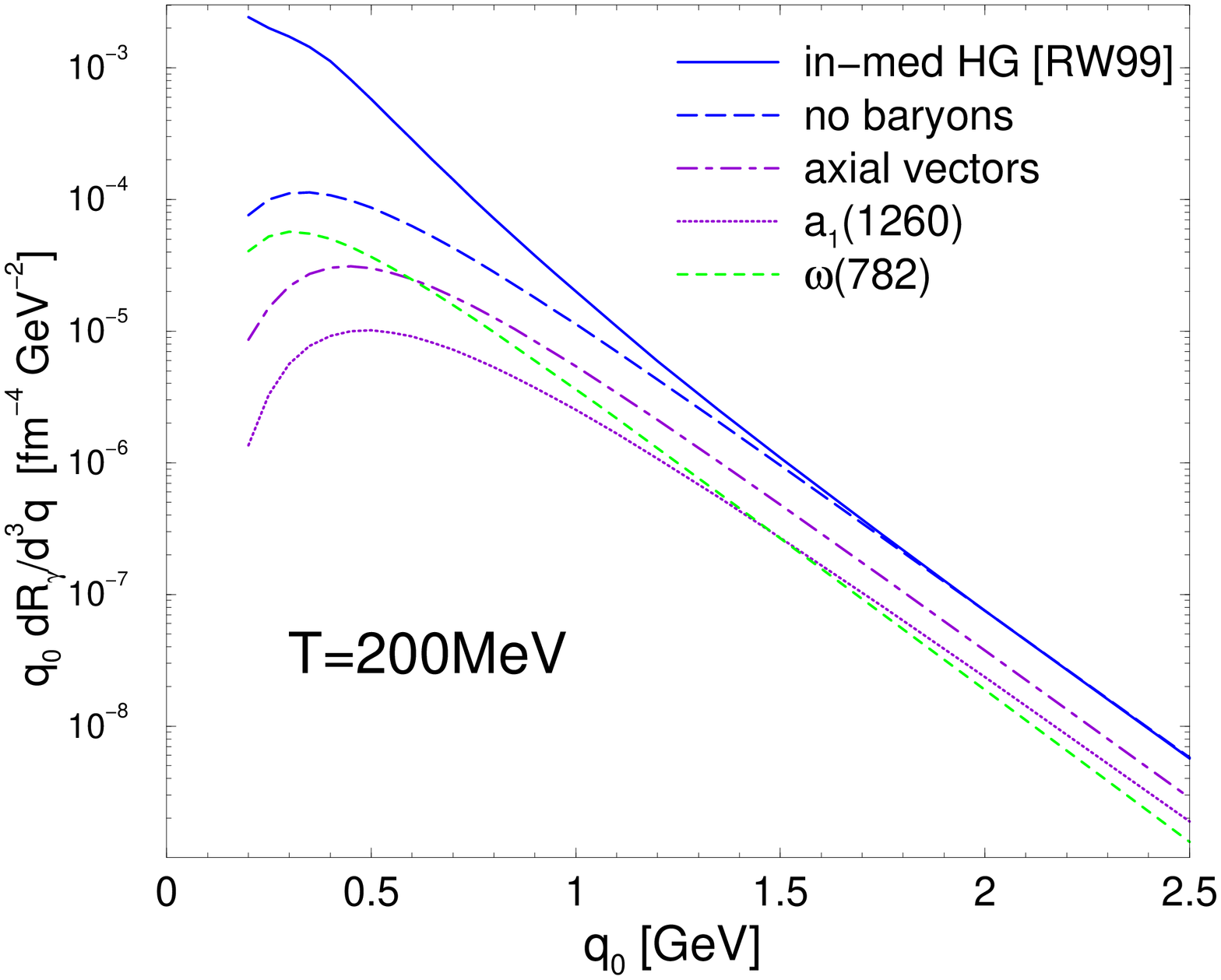,width=6.85cm,angle=-90}
\ece
\caption{(Color online) Thermal photon production rate (under conditions resembling
CERN-SPS $Pb$(158AGeV)+$Pb$ collisions) in the hadronic many-body
approach of Refs.~\protect\cite{RCW97,RW99,RG99} based on an in-medium
$\rho$ spectral function. Left panel: for temperature and baryon chemical
potential $(\mu_B,T)=(340,150)$~MeV, right panel:
$(\mu_B,T)=(220,200)$~MeV.}
\label{fig_RW}
\end{figure}

The results from the hadronic many-body approach are compiled in 
Fig.~\ref{fig_RW} for two temperature-density values characteristic
for meson-to-baryon ratios at full CERN-SPS energy (160~AGeV). 
The solid curve is the net photon spectrum obtained by taking the 
full ($\rho$-meson) spectral density to the photon point, whereas
the long-dashed curve represents the non-baryonic (sub-) component.
One observes that  the low-energy regime, $q_0\lsim 1$~GeV,
of the photon emissivity is dominated by baryonic effects (quite reminiscent
to what has been found for low-mass dileptons). These are mostly due to
direct $\rho N$ resonances such as $\Delta(1232)$, $N(1520)$, as well
as $\Delta(1232)N^{-1}$ and $NN^{-1}$ excitations in the two-pion cloud
of the $\rho$ (which, to leading order in density, correspond
to $t$-channel one-pion exchange (OPE) in processes of type
$\pi N\to \gamma N$). These contributions should be 
rather reliable for baryon densities up to at least normal nuclear
matter density, $\rho_0=0.16$~fm$^{-3}$, being constrained by
photo-absorption spectra on nucleons and nuclei~\cite{RUBW98} (including
hadronic vertex form factors with rather soft cutoff parameters 
around 600~MeV).
At comparable baryonic densities this approach yields about a factor
of two more photons than the results obtained in Ref.~\cite{SYZ97},
where only nucleonic degrees of freedom where accounted for within the 
(on-shell) chiral reduction formalism (see Ref.~\cite{SZ99} for an update
including $\Delta(1232)$ and $N(1520)$ resonances).

Beyond $\sim$~1~GeV, mesonic (resonance) states become the dominant source
of photons in the many-body approach, which includes radiative
decays of $\omega(782)$, $h_1(1170)$, $a_1(1260)$,
$f_1(1285)$, $\pi(1300)$, $a_2 (1320)$, $\omega(1420)$, $\omega(1650)$, 
$K^*(892)$ and $K_1(1270)$.
In particular, the $\omega\to\pi\gamma$ decay exhibits an appreciable
low-energy strength, consistent with the early results
of Ref.~\cite{KLS91}. Note that all hadronic vertices carry
(dipole) form factors with typical cutoff parameters of around
1~GeV, as extracted from an optimal fit to measured hadronic
and radiative branching ratios within VDM~\cite{RG99};
$t$-channel exchange processes between mesons as discussed 
in Sec.~\ref{sec_mesgas} (\eg~OPE or $a_1$-exchange
in $\pi \rho \to \pi \gamma$) are not implicit in the spectral
densities leading to the the results of Fig.~\ref{fig_RW}. 
%Their relative importance will be addressed in the following section.
They are mostly relevant at photon energies beyond 1~GeV 
and therefore do not significantly figure into bulk (low-mass) 
dilepton production, the latter being dominated by (transverse) 
momenta $q_t\lsim M$.
As mentioned above, the underlying VDM coupling to the photon   
exclusively proceeds through the $\rho(770)$ which implies that the 
strength in the pertinent e.m. correlation function beyond mass/energy 
scales of $\sim$~1GeV is no longer correctly saturated, as it is 
restricted to two-pion-type states. 
The construction of the total emission rate will be discussed in the 
following Section.

%%%%%%%%%%%%%%%%%%%%%%%%%%%%%%%%%%%%%%%%%%%%%%%%%%%%%%%%%%%%%%%%%%%%%%%
\subsection{$\omega$ $t$-Channel Exchange and Total Rate}
\label{sec_all}
%%%%%%%%%%%%%%%%%%%%%%%%%%%%%%%%%%%%%%%%%%%%%%%%%%%%%%%%%%%%%%%%%%%%%%%
Before combining the various contributions to the thermal photon rate
we investigate one more process of potential importance which is not 
present in the above and, to our knowledge, has not been addressed 
before. 
Within the $SU(2)$ flavour symmetry, the $\omega$ is a chiral singlet, 
but is known to exhibit a large coupling to $\pi\rho$ and thus, via
VDM, to $\pi\gamma$ states. Its $s$-channel decays have indeed been
calculated as early as in Ref.~\cite{KLS91} (and are included 
above), but its $t$-channel exchange in the reaction 
$\pi\rho\to\pi\gamma$ has not. We here have calculated the pertinent
contribution to the thermal emission rate using Eqs.~(\ref{rate_kin})
and (\ref{ff_av}) with the same coupling and form factor type as for 
the $s$-channel graph~\cite{RG99} (corresponding to Fig.~\ref{fig_RW}),
see below.     

In Fig.~\ref{fig_hg}  we summarise our results for the thermal photon 
emissivities from hadronic matter as evaluated in the preceding 
Sections.   
\begin{figure}
%\vspace{-1cm}
\bce
\epsfig{file=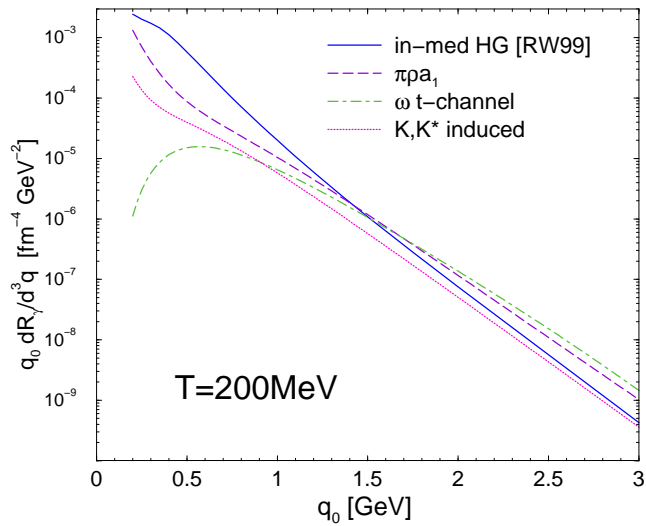,width=7cm,angle=-90}
\ece
\caption{(Color online) Compilation of thermal photon production rates from hot and 
dense hadronic matter computed in the present work at temperature 
$T$=200~MeV and baryonic chemical potential $\mu_B=$220~MeV 
(translating into total-pion to net-baryon ratio of $\sim$5). Dashed 
and dotted lines correspond to the non-/strange MYM meson gas 
emissivities of Sec.~\ref{sec_mesgas} using the parametrisations given 
in Appendix~\ref{para}, solid line to the $\rho$ spectral function  
approach including baryons, and the dashed-dotted line is solely due to 
$\omega$ $t$-channel exchange in $\pi\rho\to\pi\gamma$.}
\label{fig_hg}
\end{figure}  
At low energies, $q_0\le$~1GeV, the emission rate from the hadronic 
many-body approach ($\rho$ spectral function)~\cite{RW99}, with 
major contributions from baryonic sources, dominates. Between energies
of 1 and 2~GeV, meson gas emissivities become competitive and
eventually dominate the rates at high energies. Remarkably, the 
$\omega$ $t$-channel exchange in $\pi\rho\to\pi\gamma$ is the single
most important process beyond energies of $q_0$$\simeq$2~GeV. The
strangeness component in the production rate does not exceed 
10-15\% at any energy. 

We finally have to address the question of how to combine
the various hadronic sources, computed in two different
frameworks (cf.~Sec.~\ref{sec_def}), into the total emission rate.
Two issues arise when simply adding all of the emission rates
shown in Fig.~\ref{fig_hg}: double-counting and coherence. 
The $a_1$ $s$-channel graph is present in both $\rho$ spectral function 
and the MYM framework.
We remove it from the former, where it plays a minor role, whereas it
induces significant interference effects in the $\pi\rho a_1$ complex.
If coherence is unimportant, $t$-channel contributions can be evaluated
separately. It was verified that this was the case for the $\omega$
exchange, so that the incoherent addition of the $t$-channel
contribution is justified.

We believe that it is fair to say that the enumeration of hadronic 
photons sources given in this Section, together with form factor 
inclusions, currently represents the most realistic evaluation of the 
full hadron gas emissivity.

%%%%%%%%%%%%%%%%%%%%%%%%%%%%%%%%%%%%%%%%%%%%%%%%%%%%%%%%%%%%%%%%%%%%%%%
\subsection{Comparison to QGP Emission}
\label{sec_qgp}
\begin{figure}[!ht]
%\vspace{-1cm}
\bce
\epsfig{file=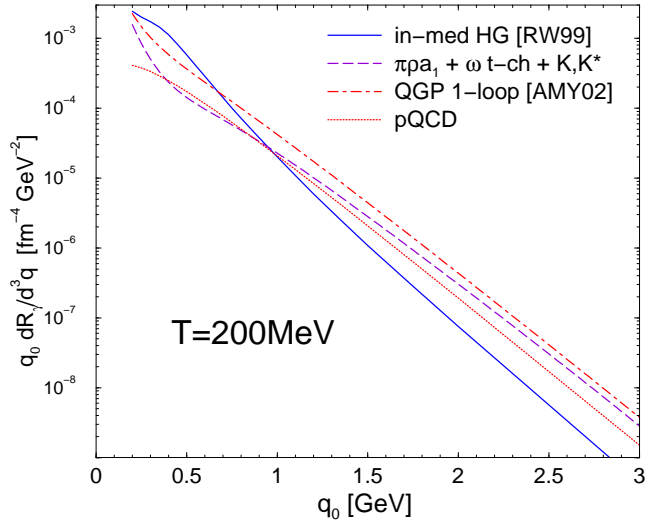,width=7cm,angle=-90}
\ece
\caption{(Color online) Comparison of HG and QGP photon production rates at $T$=200~MeV.
Solid line: hadronic many-body approach of
Refs.~\protect\cite{RCW97,RW99,RG99} (solid curve), dashed line:
mesonic contribution including hadronic form factors,
dotted line: simple pQCD result~\protect\cite{KLS91} according to
Eq.~(\ref{rate_pQ}), dashed-dotted line: complete leading-order QGP
emission~\protect\cite{AMY02}.}
\label{fig_HQ}
\end{figure}

%%%%%%%%%%%%%%%%%%%%%%%%%%%%%%%%%%%%%%%%%%%%%%%%%%%%%%%%%%%%%%%%%%%%%%%
Before turning to applications in heavy-ion reactions, our estimates for 
hadronic production rates are confronted  
with the ones from QGP emission, in particular with the simple 
lowest-order HTL-corrected pQCD result~\cite{KLS91}, 
\beq
q_0 \frac{dR_\gamma}{d^3q} = \frac{6}{9} \frac{\alpha\alpha_S}{2\pi^2}  
       T^2 {\rm e}^{-q_0/T} 
       \ln\left(1+\frac{2.912}{4\pi\alpha_s} \frac{q_0}{T} \right) \  ,
\label{rate_pQ}
\eeq
and (a parametrization of) the complete leading-order (in $\alpha \alpha_{\rm
s}$) analysis~\cite{AMY02}, cf. Fig.~\ref{fig_HQ}. 
Clearly, due to the approximations implied by each curve, none
of them can be expected to be accurate under conditions arising 
in practice, \ie~in the phase transition region. 
Nonetheless, the observation that the complete leading-order QGP calculation 
(dashed-dotted curve) is similar to the full hadronic result 
(sum of solid and long-dashed curves) 
within a factor of $\sim$ 2 over essentially all (relevant) energies
below 3~GeV might not be a mere coincidence. 
A similar behaviour has been found before for dilepton production 
rates~\cite{RW99}, perhaps suggesting a type of  ``quark-hadron 
duality'' for e.m. emission close to the expected phase boundary.
It would be most valuable to shed further light on this issue
from first principle lattice calculations, which, at the moment, 
are only reliable at sufficiently large invariant masses where 
they are, as expected, close to perturbative results~\cite{Ka02}.

%%%%%%%%%%%%%%%%%%%%%%%%%%%%%%%%%%%%%%%%%%%%%%%%%%%%%%%%%%%%%%%%%%%%%%%
\section{Photon Spectra in Ultrarelativistic Heavy-Ion Collisions}
\label{sec_urhic}
%%%%%%%%%%%%%%%%%%%%%%%%%%%%%%%%%%%%%%%%%%%%%%%%%%%%%%%%%%%%%%%%%%%%%%%

%%%%%%%%%%%%%%%%%%%%%%%%%%%%%%%%%%%%%%%%%%%%%%%%%%%%%%%%%%%%%%%%%%%%%%%
\subsection{Hard Photons and Thermal Fireball Evolution}
%%%%%%%%%%%%%%%%%%%%%%%%%%%%%%%%%%%%%%%%%%%%%%%%%%%%%%%%%%%%%%%%%%%%%%%
For a realistic comparison with direct photon spectra as extracted 
in heavy-ion collisions two further ingredients are required.

First, the thermal rates of the previous sections have to be 
convoluted  over the space-time history of the reaction. Assuming that 
thermal equilibrium can be established and maintained, hydrodynamic 
simulations are the method of choice, see \eg~Ref.~\cite{Huo02}. 
Here we employ a more simple fireball model~\cite{RW99,RS00}, which
incorporates essential elements of hydrodynamic 
calculations. The fireball evolution is started at a "formation" 
(or thermalization) time $\tau_0\le 1$~fm/c, which relates 
to the initial longitudinal 
extent of the firecylinder as $\Delta z\simeq \Delta y \tau_0$ with 
$\Delta y\simeq 1.8$ corresponding to the approximate rapidity 
coverage of a thermal distribution.    
The subsequent volume expansion, $V_{FB}(\tau)$, is carried through 
QGP, mixed and hadronic phases until "thermal" freezeout at 
$T_{fo}$=100-120~MeV, where hadrons cease to interact.  
The equations of state (EoS) for QGP and HG are modelled via thermal 
quasiparticles and a resonance gas (including about 50 species), 
respectively. Based on the conservation of net baryon-number, $N_B$, 
and total entropy, 
$S$, one is able to extract the temperature and baryon chemical 
potential at any given (proper) time, thereby defining a trajectory 
in the $\mu_B$-$T$ plane. The transition from the QGP to HG phase 
is placed at "chemical freezeout" points extracted from hadron ratios 
in experiment~\cite{pbm}. Consequently, in the HG evolution from 
chemical to thermal freezeout, hadrons stable under strong interactions 
(pions, kaons, etc.) have to be conserved explicitly by introducing 
associate chemical potentials ($\mu_\pi$, $\mu_K$, etc.). 
This has not been done in previous calculations of thermal photon
production~\cite{Dumi95,SS01,Huo02}, and induces a significantly faster
cooling in the hadronic phases~\cite{Ra02c}. In addition, at collider 
energies (RHIC and LHC), the conservation of the observed 
antibaryon-to-baryon ratio (which at midrapidities is no longer small) 
in the hadronic 
evolution becomes important~\cite{Ra02b}. An accordingly introduced
(effective) chemical potential for antibaryons has been shown to 
impact the chemistry at later stages appreciably~\cite{Ra02b} 
(in particular, it is at the origin of large meson-chemical potentials,
again implying faster cooling).   
For each collision energy, the value of the specific entropy, $S/N_B$, 
is fixed to reproduce observed hadron abundances.  
The total yield of thermal photons in an $A$-$A$ collision then
follows as
\beq
q_0 \frac{dN_\gamma^{thermal}}{d^3q}(q_t)= \frac{1}{\Delta y} 
\int\limits_{y_{min}}^{y_{max}} dy \int d\tau V_{FB}(\tau) 
 \left(q_0 \frac{dR}{d^3q} \right) \ , 
\eeq 
averaged over a rapidity interval, $[y_{min},y_{max}]$ according to 
the  experimental coverage ($\Delta y =y_{max}-y_{min}$). To incorporate 
transverse expansion in the spectra, in the thermal rest frame isotropic 
photon momentum distributions are boosted into the laboratory (lab)
frame using an average of about 70\% of the time-dependent transverse
(surface) expansion velocity at each moment in the fireball evolution.  
 
Second, an additional contribution to direct photon spectra arises
from prompt photons in primordial $N$-$N$ collisions.
The minimal baseline for a heavy-ion reaction constitutes the 
collision-number scaled expectation from proton-nucleon collisions. 
An accurate description thereof is still a matter of 
debate~\cite{Aur99}, so that we here employ empirical scaling relations 
in $x_t=2q_t/\sqrt{s}$, extracted from fits to data in 
Ref.~\cite{Sr01}. For $x_t\gsim 0.1$, corresponding to 
fixed target energies ($\sqrt{s}\lsim 50$~GeV and photon transverse 
momenta $q_t$ above 2~GeV), the cross section fit (at midrapidity, 
$y$=0) reads
\beq
q_0 \frac{d^3\sigma_\gamma^{pp}}{d^3q} = 
575 \ \frac{(\sqrt{s})^{3.3}}{(q_t)^{9.14}} \ 
\frac{{\rm pb}}{{\rm GeV^2}} \ , 
\label{dsig_pp}
\eeq
whereas for $x_t$$\lsim$0.1, corresponding to collider energies  
($\sqrt{s}\ge 200$~GeV and $q_t$$\lsim$10~GeV),
\beq
q_0 \frac{d^3\sigma_\gamma^{pp}}{d^3q} = 
6495 \ \frac{\sqrt{s}}{(q_t)^{5}} \ 
\frac{{\rm pb}}{{\rm GeV^2}} \  .  
\label{dsig_pp2}
\eeq
The naive extrapolation to a collision of two nuclei $A$ and $B$ at 
impact parameter $b$ predicts the prompt  photon spectrum to be 
\bea
q_0 \frac{dN_\gamma^{prompt}}{d^3q}(b;q_t,y=0;\sqrt{s}) &=&
q_0 \frac{d^3\sigma_\gamma^{pp}}{d^3q} \ A B T_{AB}(b) 
\nonumber\\
&=& q_0 \frac{d^3\sigma_\gamma^{pp}}{d^3q} \ 
 \frac{N_{coll}}{\sigma_{pp}^{in}} \ 
\label{dNpp}
\eea
with $T_{AB}$: nuclear overlap function, $N_{coll}$: number of 
primordial $N$-$N$ collisions, $\sigma_{pp}^{in}$: inelastic $N$-$N$ 
cross section  (we also supplement Eq.~(\ref{dNpp}) with a smooth 
cutoff for $q_0\le$~2GeV, where the parametrization (\ref{dsig_pp})
is no longer reliable).  
The lack of a consistent microscopic description of photon production 
in $p$-$p$ complicates the task to assess 
nuclear corrections, such as shadowing or intrinsic $k_T$ broadening
(Cronin effect), see \eg~Ref.~\cite{PT02} for a recent discussion. 
As a substitute for a more rigorous calculation, we here adopt the 
following strategy: since the intrinsic $k_T$ effects at the $N$-$N$ 
level are in principle contained in the parametrization, 
Eq.(\ref{dsig_pp}), the nuclear effect is approximated by fitting an 
additional (nuclear) $k_T$-smearing to $p$-$A$ data.
\begin{figure}[b!]
\bce
\epsfig{file=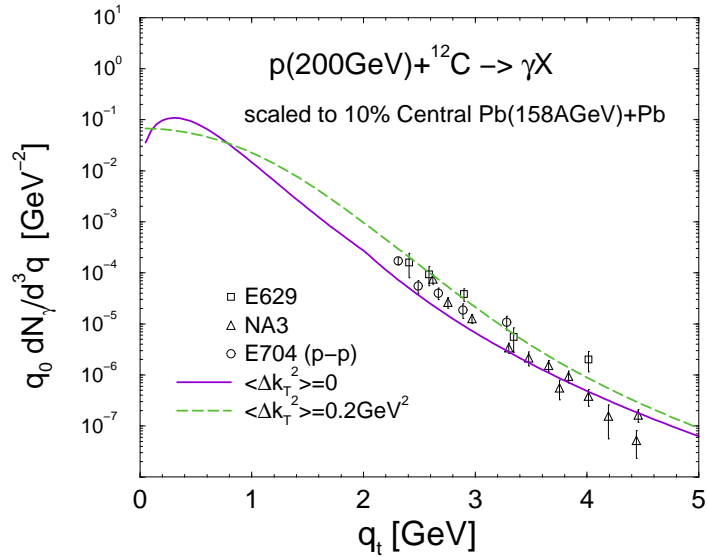,width=7.5cm,angle=-90}
\ece
\caption{(Color online) Direct photon data in proton-Carbon collisions, scaled to
central $Pb$-$Pb$ collisions at SPS energies (see text for details).
The curves show the effect of the broadening of the primordial photon
spectrum generated by the nuclear medium. The data are from
Refs.~\protect\cite{e629} (E629) and \protect\cite{na3} (NA3).}
\label{fig_cronin}
\end{figure}
The latter is modelled by folding the parameterized spectrum over a 
Gaussian distribution
\bea
f (k_T) = \frac{1}{\pi \langle \Delta k_T^2 \rangle} \ 
{\rm e}^{-k_T^2/\langle \Delta k_T^2 \rangle} \ . 
\eea
The result, together with proton-nucleus data on photon production is 
shown in Fig.~\ref{fig_cronin}. The data have been scaled to the 10\% 
central $Pb$-$Pb$ cross section at 158~AGeV according to the procedure 
used in Ref.~\cite{wa98}. 
Fitting the $p$-$A$ single photon data in this fashion, an adequate 
reproduction of the experimental measurements emerges with  
$\langle\Delta k_T^2\rangle$$\simeq$0.1-0.2~GeV$^2$.

In principle, there is a third source of photons corresponding to
emission after initial nuclear impact, but before the formation
time $\tau_0$ (the "pre-equilibrium" contribution). It is difficult to
assess both theoretically and experimentally; a rough (but
uncontrolled) estimate might be had by choosing a somewhat
smaller formation time. Note that, in principle, the modelling of those contributions 
is accessible to {\it ab-initio} simulations \cite{bms}.

%%%%%%%%%%%%%%%%%%%%%%%%%%%%%%%%%%%%%%%%%%%%%%%%%%%%%%%%%%%%%%%%%%%%%%%
\subsection{SPS}
%%%%%%%%%%%%%%%%%%%%%%%%%%%%%%%%%%%%%%%%%%%%%%%%%%%%%%%%%%%%%%%%%%%%%%%
In this section we compute transverse momentum spectra
at midrapidities from 10\% central $Pb$(158AGeV)+$Pb$ collisions for 
which photon spectra have been measured by WA98~\cite{wa98}.  
Let us first focus on thermal emission; 
QGP radiation is always calculated with the complete leading-order  
result~\cite{AMY02}, and hadronic radiation as the sum of the hadronic 
many-body~\cite{RW99,RG99}, $\pi\rho K^*K$ gas (within MYM) 
and $\omega$ $t$-channel exchange contributions with form factors
as discussed in Sec.~\ref{sec_all}.    

\begin{figure}[htb!]
%\vspace{-1cm}
\bce
\epsfig{file=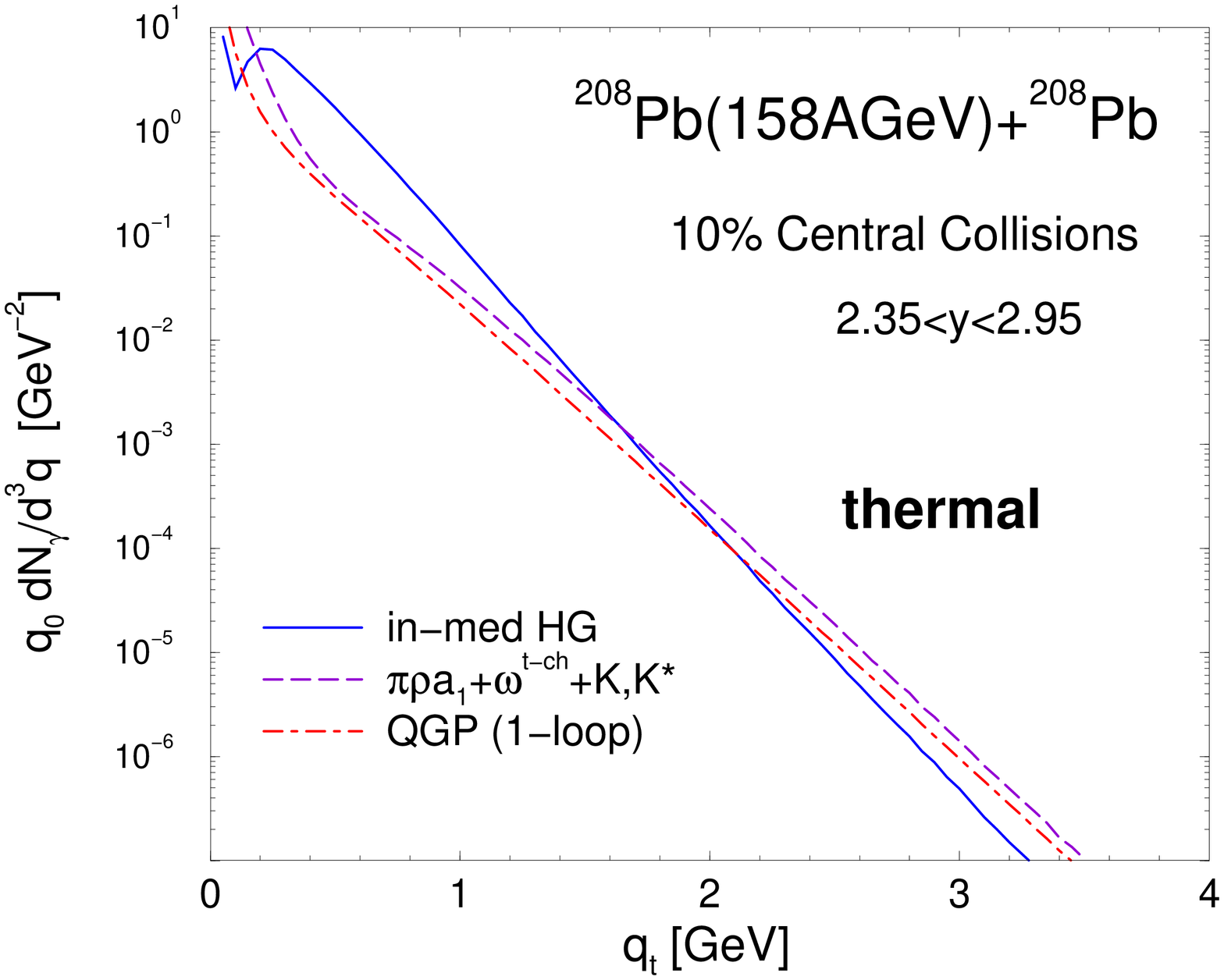,width=6.7cm,angle=-90}
\hspace{0.5cm}
\epsfig{file=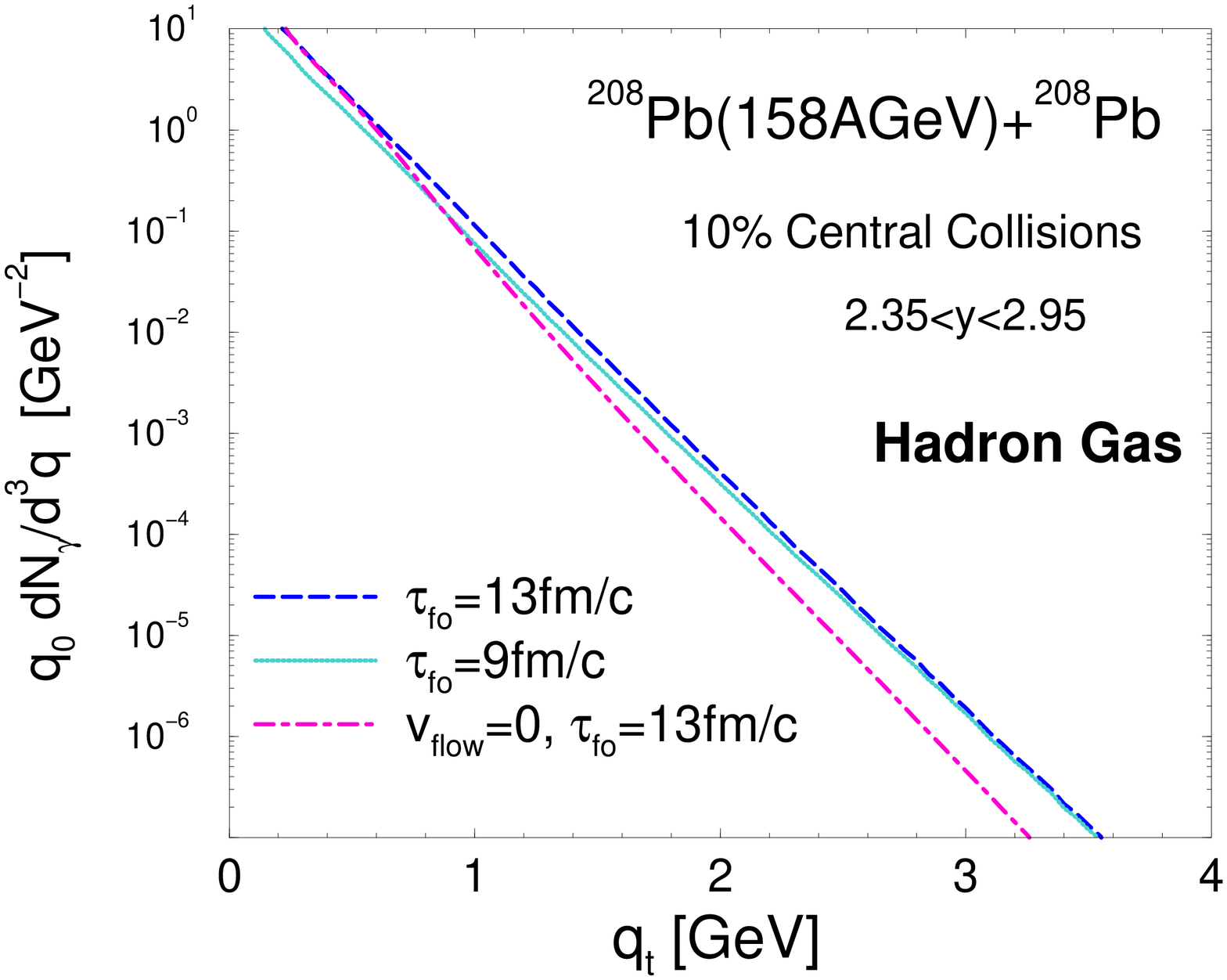,width=6.7cm,angle=-90}
\ece
\caption{(Color online) Integrated photon emission from various thermal sources in
an expanding fireball model for
central $Pb$+$Pb$ collisions at SPS. Left panel: hadronic emission
($T\le T_c$=175~MeV) from the meson gas component (dashed line) and the
in-medium $\rho$ spectral function (solid line) compared to QGP
emission ($T_c\le T\le T_i$=205~MeV) (dashed-dotted line).
Right panel: Sensitivity of the total HG yield to thermal freezeout
(long-dashed line: $\tau_{fo}$=13~fm/c corresponding to $T_{fo}$=106~MeV,
dotted line: $\tau_{fo}$=9~fm/c corresponding to $T_{fo}$=135~MeV)
and transverse flow (dashed-dotted line: $\tau_{fo}$=13~fm/c
with no transverse boost of the emission source).}
\label{fig_sps_therm}
\end{figure}

In Fig.~\ref{fig_sps_therm} we display results for a rather standard 
fireball evolution with initial $z_0$=1.8~fm (corresponding to 
$\tau_0\simeq 1$~fm/c and initial temperature $T_i=205$~MeV) and 
final temperature $T_{fo}$$\simeq 110$~MeV reached after a total 
lifetime of $\sim$~13~fm/c.
Due to transverse expansion, the total hadron gas yield outshines 
QGP emission at all momenta. This is in close reminiscence to the 
calculations of intermediate dilepton spectra within the same 
framework~\cite{RS00}, where QGP radiation was found to constitute 
about 30\% of the thermal component that was able to reproduce 
the excess observed by the NA50 collaboration~\cite{na50-int}    
(see also Ref.~\cite{kvas}). 
As expected, photons of baryonic origin prevail in the spectrum 
for $q_t$$\lsim$1~GeV; this region is thus intimately related 
to the low-mass (and low-transverse momentum) dilepton enhancement 
observed by CERES/NA45~\cite{ceres160}. The same feature 
has been found~\cite{RW00-2} when comparing the hadronic 
many-body contributions to the upper limits  
in $S$(200AGeV)+$Au$ by WA80~\cite{wa80}.   
In the right panel of Fig.~\ref{fig_sps_therm} we illustrate the 
sensitivity of the hadronic thermal emission to properties of the
fireball evolution. When increasing the thermal freezeout time
from our (SPS) default value of 106~MeV to 135~MeV, the yield
at $q_t$$\le$1GeV is reduced by up to 30\%, whereas it is 
essentially unchanged beyond $q_t$$\simeq$2~GeV, thus reflecting
emission close to $T_c$. On the contrary,
if the additional boost on the photons due to the transverse expansion
is neglected, the high-momentum spectrum is reduced appreciably (by
a factor of $\sim$3 already at $q_t$=2~GeV), whereas the low-momentum 
region is only mildly affected.  

\begin{figure}[tbh!]
%\vspace{-1cm}
\bce
\epsfig{file=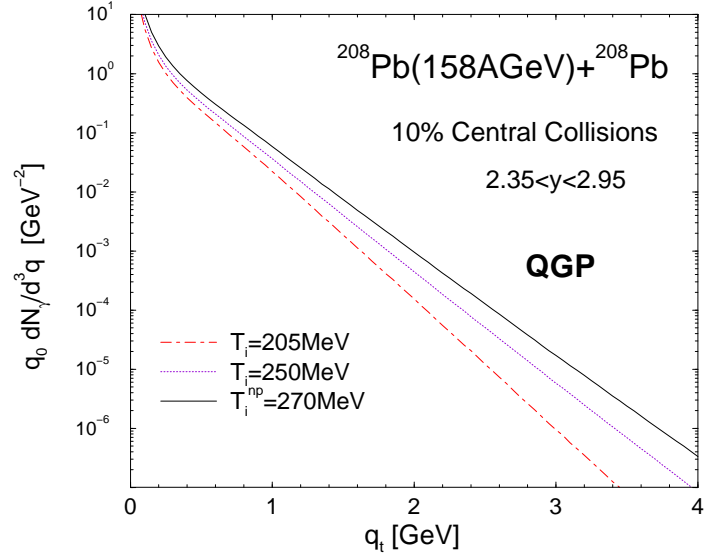,width=7.5cm,angle=-90}
\ece
\caption{(Color online) Sensitivity of the QGP photon emission yield within an
expanding fireball model for central $Pb$+$Pb$ collisions at SPS.}
\label{fig_sps_qgp}
\end{figure}

As is well-known, high-energy (-mass) photon (dilepton) emission 
is rather sensitive to initial temperatures in heavy-ion reactions,
due to the large (negative) exponents in the thermal factors.  
This is confirmed by our results for the QGP contribution in 
Fig.~\ref{fig_sps_qgp} when decreasing the formation time 
from 1 to 0.56~fm/c, the latter implying $T_i$=250~MeV. 
Another effect that has been ignored in available hydrodynamic 
calculations so far is associated with corrections to the 
QGP equation of state. The standard assumption is that of 
an ideal (massless) gas with an effective number of flavours
$N_f=2.5$ corresponding to a total degeneracy 
$d_{QGP}$=(10.5$N_f$+16)=42 (in our default calculations 
we use $d_{QGP}$=40). However, lattice gauge theory results~\cite{KLP00} 
indicate $\sim$20\% smaller values than the ideal gas 
for the thermodynamic state variables in the for SPS energies 
relevant temperature region $T$=1-2~$T_c$. Implementing such 
a reduction into the entropy density (which is the relevant quantity
for the fireball evolution) by using $d_{QGP}$=32 yields an increase 
of the initial temperature from 250~MeV to 270~MeV  without 
decrease in initial volume~\cite{Ra02c,Renk:2003fn}. 
The resulting thermal photon spectrum from QGP radiation shows 
appreciable sensitivity to this modification at high energies, 
cf.~solid vs. dotted curve in  
Fig.~\ref{fig_sps_qgp}. This sensitivity does not persist 
into the low-energy region $q_0\le 1$~GeV, and thus does not affect
low-mass dilepton production~\cite{Ra02a}.     

Let us now turn to a comparison with the recent measurements of
WA98~\cite{wa98}. 
\begin{figure}[bht!]
%\vspace{-1cm}
\bce
\epsfig{file=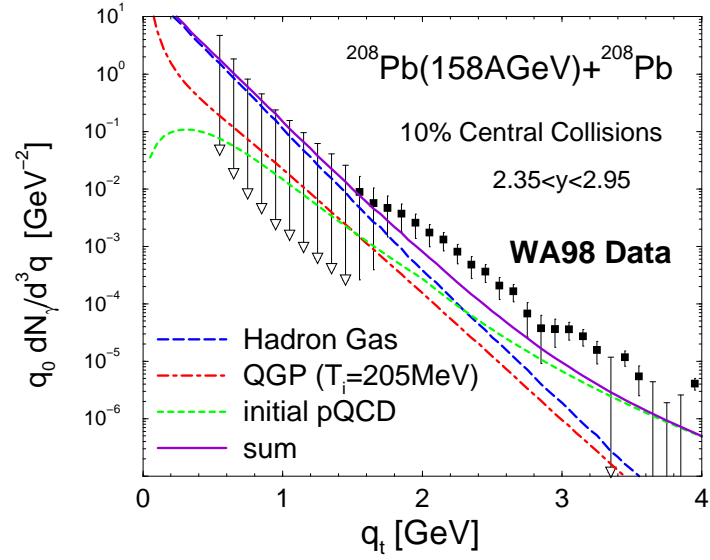,width=7.5cm,angle=-90}
\ece
\caption{(Color online) Thermal plus prompt photon spectra compared to
data from WA98~\protect\cite{wa98}
for central $Pb$+$Pb$ collisions at SPS .}
\label{fig_ph_all}
\end{figure}
Our baseline scenario
consists of thermal emission (hadronic and QGP) from the expanding
fireball with $T_i$=205~MeV, supplemented by prompt (pQCD) photons
from primordial $N$-$N$ collisions without any nuclear effects,
cf. Fig.~\ref{fig_ph_all}.
Up to transverse momenta of about 1.5~GeV the data (upper limits)
are essentially saturated by thermal radiation from the hadronic
phase.
This is gratifying to note since this regime, as discussed
above, is directly related to the low-mass dilepton excess observed
by CERES/NA45~\cite{ceres160,ceres40}, which can be successfully 
described within the same approach~\cite{RW99}.   
Beyond 3~GeV, prompt photons dominate, but do not seem
to provide enough yield to account for the data.
Since the hadron gas emission is essentially fixed and describes
well the low-energy regime, three possibilities are left for the
origin of discrepancies above $q_0$$\simeq$2~GeV:
(i) modifications of the prompt yield, 
(ii) pre-equilibrium emission,  (iii) larger QGP radiation.
In the following, cases (i) and (iii) (or a combination thereof)
will be investigated.  

\begin{figure}[htb!]
%\vspace{-1cm}
\bce
\epsfig{file=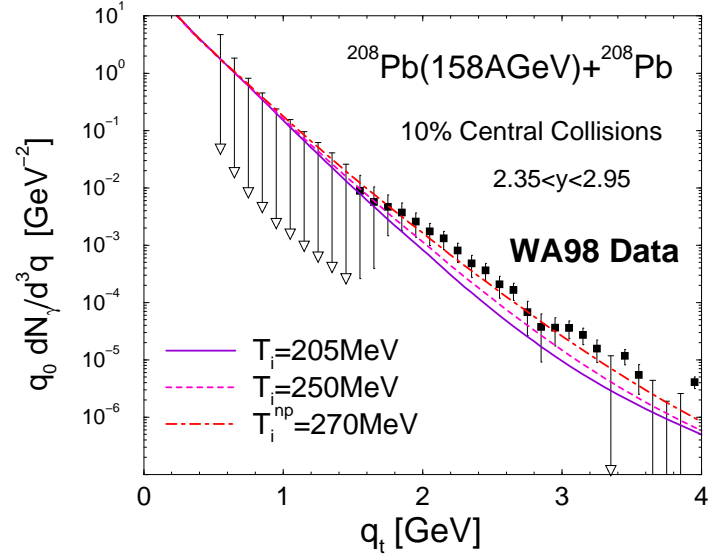,width=7.5cm,angle=-90}
\ece
\caption{(Color online) Effects of various initial temperatures on the total photon
spectra in $Pb$+$Pb$ collisions at SPS, compared to data from
WA98~\protect\cite{wa98}.}
\label{fig_ph_init}
\end{figure}

First, we study the effects of the initial temperature on the photon 
spectrum.  The exercise in the right panel of Fig.~\ref{fig_sps_therm} 
is repeated adding all sources discussed here, 
cf.~Fig.~\ref{fig_ph_init}.
Clearly, when going to (for SPS conditions) rather short formation
times of $\tau_0\simeq 0.5$~fm/c, coupled with non-perturbative 
(suppression) effects in the QGP EoS, a rather good reproduction 
of the entire spectrum can be achieved. This statement agrees with 
the hydrodynamic analyses of Refs.~\cite{SS01,Huo02}.     

\begin{figure}[!htb]
%\vspace{-1cm}
\bce
\epsfig{file=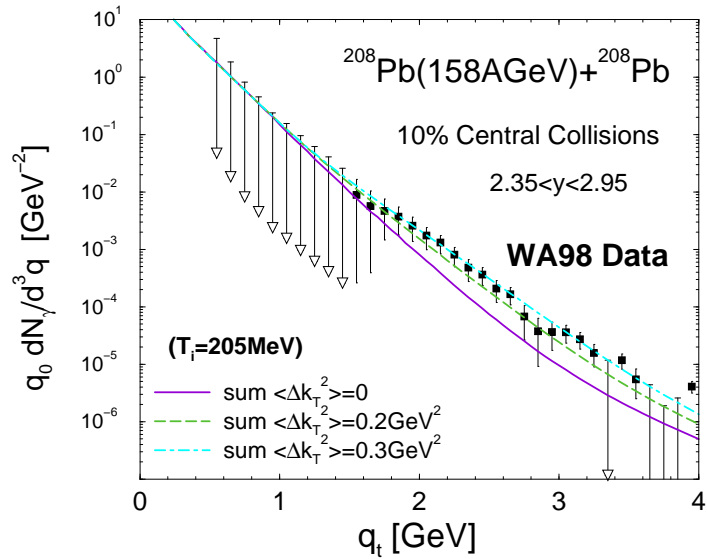,width=7.5cm,angle=-90}
\ece
\caption{(Color online) Effects of the nuclear broadening of the primordial photon
spectrum on the measured spectrum. All the sources discussed in this
paper are included in the space-time evolution.}
\label{fig_sps_cron}
\end{figure}

The second possibility relates to the nuclear Cronin enhancement, which
we implement as outlined in the previous Section. 
The usual assumption to extrapolate nuclear broadening effects on 
\eg~$\pi^0$ or $\gamma$ spectra is that
\beq
\langle \Delta k_T^2 \rangle_{AA} = N \langle \Delta k_T^2 \rangle_{pA} \ ,
\eeq
with $N$=2~\cite{Dumi01}. Alternatively, based on a careful analysis
of the target ($A$) dependence in $p$-$A$ collisions, it has been
suggested in Ref.~\cite{Papp99} that the Cronin effect is due to no more
than one {\em semi}-hard collision prior to the hard scattering, and
therefore saturating as a function of the $N$-$N$ collision number. In 
this eventuality, $N \leq$ 2. Recalling that, from Fig.~\ref{fig_cronin}, 
$\langle \Delta k_T^2\rangle $=0.1-0.2~GeV$^2$ gives a reasonable description of the 
$\gamma$ spectra in $p$-$C$, $\langle \Delta k_T^2 \rangle$-values between 0.2 and 
0.3~GeV$^2$ seem appropriate for central $Pb$-$Pb$ collisions. One 
should also note that the pertinent spectral enhancement in the 
$q_t$$\simeq$3~GeV region amounts to a factor of around 3, which is 
quite consistent with the nuclear enhancement in $\pi^0$ production 
observed in the same experiment~\cite{wa98_pi0}.
In Fig.~\ref{fig_sps_cron} we have combined the baseline thermal yield 
($\tau_0$=1~fm/c, \ie~$T_i$=205~MeV) with 3 values for the nuclear 
$k_T$-broadening, \ie~$\langle \Delta k_T^2 \rangle$=0, 0.2 and 0.3~GeV$^2$.  
The thermal plus Cronin-enhanced pQCD spectra provide good description 
of the WA98 data, even with an initial temperature as low as 
$T_i$=205~MeV. This constitutes one of the main results of our work: 
the photon spectrum in nucleus-nucleus collisions at SPS energies is 
perfectly compatible with ``moderate'' initial temperatures.
It also complements, within a common thermal framework, earlier 
descriptions of low- and intermediate-mass dilepton 
spectra~\cite{RW99,RS00}, as well as $J/\psi$ and $\psi'$ production
systematics~\cite{GR02}, as observed by the CERN-SPS
experiments CERES/NA45~\cite{ceres160,ceres40} and 
NA50~\cite{na50-int,na50-psi}, respectively.    

It is also of interest to quote the values of the transverse momentum
where the pQCD yield exceeds the total thermal one; these
are $q_t$=2.55, 1.7 and 1.55~GeV, corresponding to 
$\langle \Delta k_T^2 \rangle$=0, 0.2 and 0.3~GeV$^2$, respectively. Again, this
compares well with the calculation of intermediate-mass dileptons in 
Ref.~\cite{RS00}, where the Drell-Yan contribution was
found to exceed the thermal one at $M_{\mu\mu}$$\simeq$2~GeV (note
that the Cronin effect is expected to be less pronounced for
quarks than for gluons, and thus will affect the Drell-Yan process less 
than prompt photons).

%%%%%%%%%%%%%%%%%%%%%%%%%%%%%%%%%%%%%%%%%%%%%%%%%%%%%%%%%%%%%%%%%%%%%%%
\subsection{RHIC and LHC}
\label{sec_rhic}
%%%%%%%%%%%%%%%%%%%%%%%%%%%%%%%%%%%%%%%%%%%%%%%%%%%%%%%%%%%%%%%%%%%%%%%
At collider energies the space-time evolution of the expanding 
QGP and hadronic fireball is expected to change in several respects. 
First, higher energies entail larger charged 
particle multiplicities per unit rapidity, $dN_{ch}/dy$. In central 
$Au$+$Au$ collisions at full RHIC energy ($\sqrt{s}=200$~AGeV) 
experiments have found~\cite{brahms,phobos} about 
a factor of 2 increase as compared to maximum SPS energy  
($\sqrt{s}=17.3$~AGeV). Extrapolations into the LHC regime  
($\sqrt{s}=5500$~AGeV) suggest another factor of up to $\sim$4 
enhancement over the RHIC results.  

Second, the {\em net} baryon content at midrapidity decreases, 
implying small baryon chemical potentials at chemical freezeout,
\eg~$\mu_B\simeq 25$~MeV at RHIC-200. At the same time, the observed 
production of baryon-antibaryon pairs strongly rises, resulting in 
{\em total} rapidity densities for baryons at RHIC that are quite 
reminiscent of the situation at SPS energies~\cite{Tse02}. 
This observation not only necessitates the explicit conservation of 
antibaryon-number between chemical and thermal
freezeout~\cite{Ra02b} (see above), but also requires to evaluate
baryonic photon sources with the {\em sum} of the baryon and
antibaryon density (strong and e.m.~interactions are CP-invariant).   

Third, the transverse expansion (\ie~flow velocity) increases 
by about 20\% from SPS to RHIC (presumably further at LHC), whereas the 
total fireball lifetime does not appear to change much. The latter, 
however, is likely to increase at LHC, due the significantly 
larger system sizes towards thermal freezeout.  

All these features are readily implemented~\cite{Ra01,Ra02b} into the 
thermal fireball description employed for SPS energies above. In 
addition, the primordial pQCD component changes its $x_t$-scaling
behavior~\cite{Sr01} which is accounted for by replacing the 
parametrization Eq.~(\ref{dsig_pp}) by Eq.~(\ref{dsig_pp2}). For 
simplicity we here refrain from introducing a nuclear $k_T$ broadening, 
which is expected to be much less pronounced (and/or compensated by
shadowing corrections) at collider energies. First data on 
high-$p_t$ hadron production in $d$-$Au$ collisions at 
$\sqrt{s}$=200~GeV~\cite{phenix03,star03,phobos03} indeed indicate 
only a comparatively small enhancement of around 20-30\% over the 
spectra measured in $p$-$p$ collisions.

\begin{figure}[htb!]
\bce
\epsfig{file=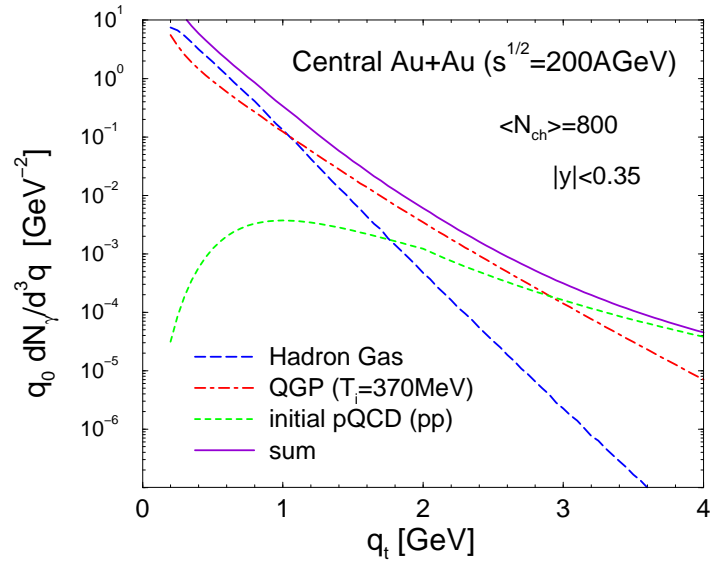,width=7.5cm,angle=-90}
\ece
\caption{(Color online) Integrated photon emission spectra from central
$Au$+$Au$ collisions at RHIC.
Short-dashed line: pQCD photons from primordial $N$-$N$ 
collisions; dashed-dotted line: thermal QGP radiation; 
long-dashed line: thermal hadron gas emission; solid line: total
direct photon yield.} 
\label{fig_rhic}
\end{figure}

Our photon predictions for full RHIC energy are summarized in
Fig.~\ref{fig_rhic}. 
The thermal component has been evaluated 
with a typical formation time, $\tau_0=1/3$~fm/c, as used before 
in dilepton~\cite{Ra01} and charmonium~\cite{GR02} applications
(it is also consistent with hydrodynamic approaches that correctly
reproduce the elliptic flow measurements which are particularly
sensitive to the early phases, see Ref.~\cite{KH03} for a recent 
review).  One notes that the spectrum decomposes into essentially
3 regimes: at low energies, $q_0$$\le$1~GeV, the major source 
are still thermal hadrons, whereas at high energies, $q_0\ge$~3GeV,
prompt pQCD photons dominate. The intermediate region, 
1$\le$$q_0$$\le$3~GeV, appears to be a promising window to be sensitive
for thermal QGP radiation. The latter has been calculated assuming 
chemically equilibrated quark- and gluon-densities throughout. It is
conceivable, however, that the early QGP phases are gluon-dominated, 
\ie~with quark fugacities much smaller than one (even the gluon
fugacities could be reduced). In this case, on the one hand, the photon 
emissivities at given temperature are severely suppressed. On the other
hand, if most of the total entropy is produced sufficiently early, 
smaller fugacities imply larger temperatures, thus increasing the
photon yield. The interplay of these effects has been studied for 
dilepton production in Ref.~\cite{Ra01}, where it has been found that 
the net effect consists of a slight hardening of the QGP emission 
spectrum with a pivot point at $M$$\simeq$3~GeV. For photons the 
situation might be even more favorable due to the participation of 
gluons in their production (\eg~$g+q\to g\gamma$, as opposed to 
leading-order $q\bar q\to ee$ for dileptons).  
 
\begin{figure}[ht!]
\bce
\epsfig{file=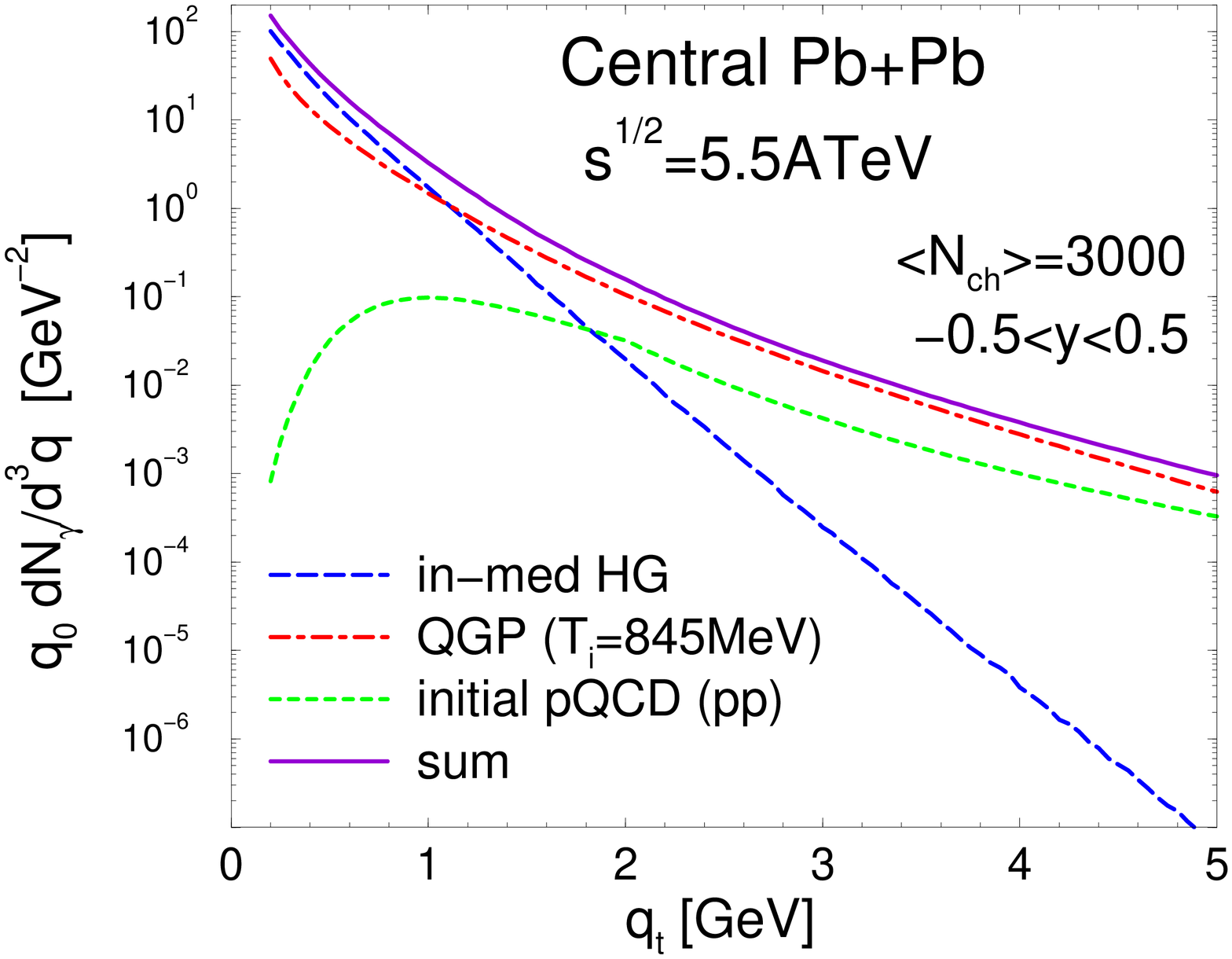,width=7cm,angle=-90}
\hspace{0.3cm}
\epsfig{file=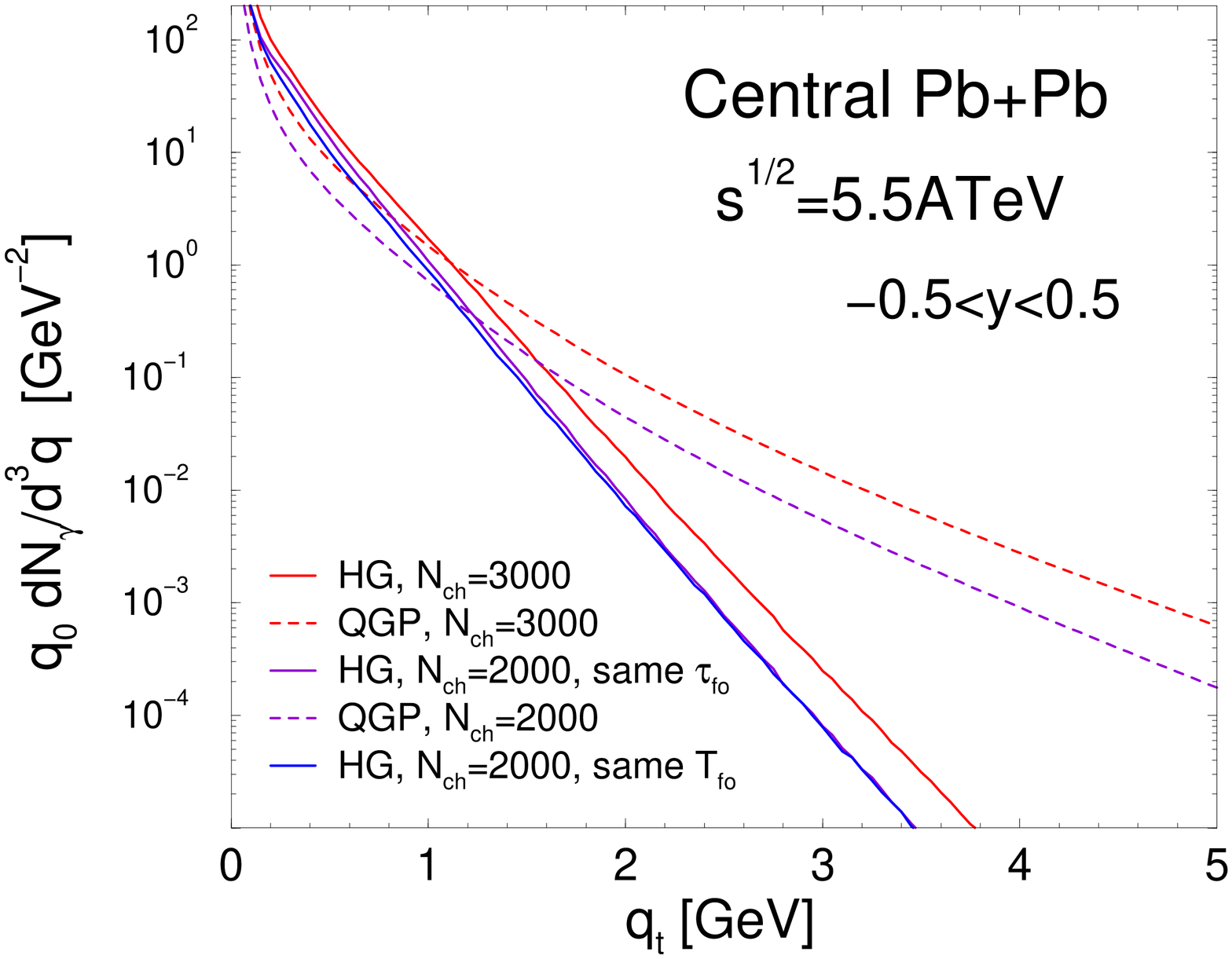,width=7cm,angle=-90}
\ece
\caption{(Color online) Left panel: integrated photon emission from 
various thermal sources in central $Pb$+$Pb$ collisions at the LHC;
line identification as in Fig.~\protect\ref{fig_lhc}. Right panel:
sensitivity of thermal emission spectra from the hadron gas (solid
lines) and the QGP (dahed lines) to the charged particle rapidity
density. For HG emission at $N_{ch}$=2000, the upper line corresponds
to assuming the same freezeout time ($\tau_{fo}$$\simeq$25~fm/c) as for 
$N_{ch}$=3000, whereas the lower line corresponds to assuming the same 
freezeout temperature ($T_{fo}$$\simeq$90~MeV) (implying a 10\% reduced 
lifetime).} 
\label{fig_lhc}
\end{figure}
We finally turn to the LHC, cf.~Fig.~\ref{fig_lhc}. 
According to our estimates, assuming a formation time of 0.11~fm/c
(translating into $T_i$$\simeq$850~MeV for $dN_{ch}/dy$$\simeq$3000),
the QGP window extends significantly further in transverse momentum 
than under RHIC conditions, cf.~left panel of Fig.~\ref{fig_lhc}, 
although this feature is sensitive to: (i) the formation time 
(ii) a possible chemical undersaturation of the QGP, (iii) nuclear
effects on the initial pQCD yield. The transition from HG to QGP 
dominated emission occurs again close to $q_t$=1~GeV. 
In the right panel of Fig.~\ref{fig_lhc} we illustrate the 
sensitivity of the thermal spectra with respect to the produced 
charged particle muliplicity 
within our schematic fireball evolution model. For simplicity, we 
assumed the same formation time and expansion parameters for both 
$N_{ch}$=3000 and 2000. We then find that the total integrated photon 
yield (\ie, for transverse momenta above 50~MeV) scales as 
$N_{ch}^\alpha$ with $\alpha$$\simeq$1.4 if the same thermal freezeout 
temperature $T_{fo}$$\simeq$90~MeV is imposed. This value for $\alpha$ 
is somewhat larger than the 1.2 found in hydrodynamic caclulations 
of Ref.~\cite{CRS98}, but confirms the deviation from the naive
quadratic behavior ($\alpha$=2). However, the latter is approached
(and even exceeded) for the yield at higher transverse momenta; \eg, 
when integrating over $q_t$ with a lower bound of 1~GeV (2~GeV), 
$\alpha$ increases to $\sim$1.9 (2.3).    

%%%%%%%%%%%%%%%%%%%%%%%%%%%%%%%%%%%%%%%%%%%%%%%%%%%%%%%%%%%%%%%%%%%%%%%%%
\section{Summary and Conclusions}
\label{sec_concl}
%%%%%%%%%%%%%%%%%%%%%%%%%%%%%%%%%%%%%%%%%%%%%%%%%%%%%%%%%%%%%%%%%%%%%%%%%

In the present article we have attempted an improved evaluation of 
hadronic thermal emission rates for real photons, suitable for 
realistic applications in relativistic heavy-ion collisions. In what we 
think is a more complete treatment than has been achieved before, our 
main findings are: 
\begin{itemize}
\item[(i)] Revisited meson gas emissivities built upon an effective 
Lagrangian of the massive Yang-Mills type lead to about 40\% reduced 
rates in a $\pi\rho a_1$ gas as compared to previous analyses. An 
inclusion of strangeness-bearing channels has revealed that the latter 
contribute at the 20\% level. A quantitative evaluation of hadronic form 
factors has been performed throughout, mandatory for applications.    
\item[(ii)]  Photon rates from the baryonic sector have been obtained 
from a limiting procedure where in-medium $\rho$ spectral densities 
were carried to the photon point. This procedure makes consistent the 
real photon analysis with that of low invariant mass dileptons, thereby 
elucidating the role of baryons in photon emission during nuclear 
collisions. Their contributions have been
shown to be substantial for photon energies $q_0$$\le$1~GeV.    
\item[(iii)] As the single most important process at high energies
we have identified $\omega$ $t$-channel exchange in the 
$\pi\rho\to\pi\gamma$ reaction, which had not been considered before.
\end{itemize}

The total hadronic emissivity has been compared to a recent 
complete leading-order (in strong and electromagnetic couplings)
QCD calculation for the QGP. In the vicinity of the expected phase 
boundary both rates turned out to be very similar, at all energies
of practical relevance.  
  
The net rates have been folded over a fireball evolution of nuclear
collisions. This approach, albeit schematic, is consistent with observed 
hadrochemistry and hydrodynamic expansion characteristics, as well as 
dilepton and charmonium data measured at SPS energies. 
Using a comprehensive fit of photon cross sections in $p$-$p$ and 
$p$-$\bar{p}$ interactions, an estimate of the Cronin effect (nuclear
$k_T$-broadening) in $p$-$A$ collisions was first extracted, then 
generalised to central $Pb$-$Pb$ collisions to address the WA98 photon 
measurements at the SPS. Combining the complete set of hadronic rates 
with QGP emission and our Cronin-effect estimates on the primordial 
photon component, we are able to reproduce the WA98 data, with moderate 
values of initial temperature ($T_i$$\simeq$200-240~MeV) and transverse
momentum broadening ($\langle \Delta k_T^2 \rangle$$\simeq$0.2-0.3~GeV$^2$).  
Predictions for photon spectra to be measured at higher colliding energies 
(RHIC and LHC) have also been made, suggesting transverse-momentum
windows around 3~GeV as promising to track QGP radiation.

%\vskip1cm
 
\begin{acknowledgments}
We are grateful to K. L. Haglin for useful communications in connection with his
parallel investigation of reactions with strangeness, discussed in 
Section~\ref{sec_mesgas} of this article. 
We are happy to acknowledge useful discussions with T. C.~Awes 
and D. K.~Srivastava. 
The work of ST and CG was supported in part by 
the Natural Sciences and Engineering Research Council of Canada, and in part 
by the Fonds Nature et Technologies of Quebec. 

\end{acknowledgments}

%%%%%%%%%%%%%%%%%%%%%%%%%%%%%%%%%%%%%%%%%%%% 

%
%
%

\newpage
\appendix

\section{Parameterisations}
\label{para}

The photon emission rates have been calculated from the Lagrangian describe in 
Sec.~\ref{sec_had} and by the VMD interaction 
\beq
{\cal L}_{em}=-C m^{2}_{\rho}A^{\mu}\rho^{0}_{\mu}
\eeq
where $A^{\mu}$ is the photon field and $C$ is a constant adjusted by the 
experimental decay $\rho^{0}\rightarrow e^{+}e^{-}$, which gives C=0.059.
In order to respect the Ward Identity in a direct way, we multiply each Feynman 
amplitude  by the square of the averaged space-like form factor of 
Eq.~(\ref{ff_t}). Time-like form 
factors have been defined to be normalised to one for on-shell decays.
We quote below parametrisations which include the axial meson $a_{1}$ as 
exchange particle for non-strange initial states. In the following, the photon 
energy ($E$) and the temperature ($T$) are both in GeV.  
Parameterisations for $K^*\rightarrow K+\pi+\gamma$ and $K+K\rightarrow 
\rho+\gamma$ do not appear because their rates have been found to be negligible.
\bea
E\frac{dR_{\pi + \rho \rightarrow \pi+\gamma}}{d^{3}p} &&= F^{4}(E)\, T^{2.8}exp\left(\frac{-(1.461 T^{2.3094}+0.727)}{(2 T E)^{0.86}}+(0.566 T^{1.4094}-0.9957)\frac{E}{T}\right)\ (\mbox{fm}^{-4}\mbox{GeV}^{-2})
\eea

\bea
E\frac{dR_{\pi + \pi \rightarrow \rho+\gamma}}{d^{3}p} &&=F^{4}(E)\, \frac{1}{T^{5}}exp\left(-(9.314T^{-0.584}-5.328)(2TE)^{0.088}+(0.3189T^{0.721}-0.8998)\frac{E}{T}\right)
\eea

\bea
E\frac{dR_{\rho \rightarrow \pi +\pi +\gamma}}{d^{3}p} &&=F^{4}(E)\, \frac{1}{T^2}exp\left(-\frac{(-35.459T^{1.126}+18.827)}{(2TE)^{(-1.44T^{0.142}+0.9996)}}-1.21\frac{E}{T}\right)
\eea

\bea
E\frac{dR_{\pi + K^{*} \rightarrow K+\gamma}}{d^{3}p} &&=F^{4}(E)\, T^{3.75}exp\left(-\frac{0.35}{(2TE)^{1.05}}+(2.3894T^{0.03435}-3.222)\frac{E}{T}\right)
\eea

\bea
E\frac{dR_{\pi + K \rightarrow K^{*}+\gamma}}{d^{3}p} &&=F^{4}(E)\, \frac{1}{T^3}exp\left(-(5.4018T^{-0.6864}-1.51)(2TE)^{0.07}-0.91\frac{E}{T}\right)
\eea

\bea
E\frac{dR_{\rho +K \rightarrow K +\gamma}}{d^{3}p} &&=F^{4}(E)\, T^{3.5}exp\left(-\frac{(0.9386T^{1.551}+0.634)}{(2TE)^{1.01}}+(0.568T^{0.5397}-1.164)\frac{E}{T}\right)
\eea

\bea
E\frac{dR_{K^* +K \rightarrow \pi +\gamma}}{d^{3}p} &&=F^{4}(E)\, T^{3.7}exp\left(\frac{-(6.096T^{1.889}+1.0299)}{(2TE)^{(-1.613T^{2.162}+0.975)}}-0.96\frac{E}{T}\right)
\eea

$F(E)$ is the form factor, cf. Sec.~\ref{sec_mesgas}.

\end{document}